\documentclass[%
 reprint,
 amsmath,amssymb,
 aps,
]{revtex4-1}

\usepackage{graphicx}% Include figure files
\usepackage{dcolumn}% Align table columns on decimal point
\usepackage{bm}% bold math
\usepackage{mathptmx}
\usepackage{amsmath,amssymb}
\usepackage{bm}
\usepackage{hyperref}
\usepackage{graphicx}
\usepackage{color}
\usepackage{float}
\usepackage{dsfont}
\usepackage{epsfig}
\usepackage{ mathrsfs }
\usepackage[hypcap]{caption}
\usepackage{graphicx}
\usepackage{caption}
\usepackage{subcaption}
\usepackage{empheq}
\usepackage{url}
\usepackage{textcomp}
\usepackage{float}
\usepackage[utf8]{inputenc}
\usepackage{amsthm}
\usepackage{framed}
\usepackage{ amssymb }
\usepackage{pgfplots}
\usepackage{tikz}
\usepackage{color}

\newcommand{\red}[1]{{\color{red} #1}}
\newcommand{\blue}[1]{{\color{blue} #1}}
\newcommand{\green}[1]{{\color{teal} #1}}

\DeclareMathAlphabet{\mathcal}{OMS}{cmsy}{m}{n}

\newcommand{\lip}{\langle}
\newcommand{\rip}{\rangle}

\newcommand{\kett}[1]{\mathop{| #1 \rangle}\nolimits}
\newcommand{\ket}[1]{\mathop{\left| #1 \right>}\nolimits}
\newcommand{\bra}[1]{\mathop{\left<#1\,\right|}\nolimits}

\newcommand{\ketbra}[2]{| #1 \rip\lip #2 |}

\newcommand{\dthresh}{\mathcal{T}_{%\text{thresh},
k}}

\newcommand{\ispa}[3]{{\vert \psi_{#1 #2}^{(#3)}\rangle }}

\newcommand{\kbihpa}[3]{{\vert \phi_{#1 #2}^{(#3)}\rangle \langle  \phi_{#1 #2}^{(#3)}\vert }}

\newcommand{\nsq}[1]{{\left\| #1 \right\|^2}}
\newcommand{\fl}[1]{{\left\lfloor #1 \right\rfloor}}
\newcommand{\cl}[1]{{\left\lceil #1 \right\rceil}}

\def\Mod{\text{ mod }}
\def\tI{\mathrm{I}}
\def\tO{\mathrm{O}}

\def\tH{\mathrm{H}}

\def\tE{\mathrm{E}}

\newcommand{\abs}[1]{\left\vert #1 \right\vert}
\newcommand{\abss}[1]{{\vert #1 \vert}}
\newcommand{\bk}[2]{\mathop{\left< #1 , #2 \right>}}
\newcommand{\bkk}[2]{\mathop{\langle #1, #2 \rangle}}

\newcommand{\mumin}{{\mu_{\text{min}}}}
\newcommand{\mumax}{{\mu_{\text{max}}}}

\newtheorem{lemma}{Lemma}
\newtheorem{proposition}{Proposition}

\newtheorem{remark}{Remark}

\begin{document}

%\preprint{APS/123-QED}

\title{Families of Quantum Fingerprinting Protocols}% Force line
\author{Benjamin Lovitz}
\email{benjamin.lovitz@gmail.com}
\affiliation{Institute for Quantum Computing, Department of Physics and Astronomy, University of Waterloo, 200 University Ave W, Waterloo, ON, Canada, N2L 3G1.}
\author{Norbert L\"utkenhaus}
\email{lutkenhaus.office@uwaterloo.ca}
\affiliation{Institute for Quantum Computing, Department of Physics and Astronomy, University of Waterloo, 200 University Ave W, Waterloo, ON, Canada, N2L 3G1, and Perimeter Institute for Theoretical Physics, 31 Caroline St N, Waterloo, ON, Canada, N2L 2Y5.}

\date{\today}% It is always \today, today,

\begin{abstract}
We introduce several families of quantum fingerprinting protocols to evaluate the equality function on two $n$-bit strings in the simultaneous message passing model. The original quantum fingerprinting protocol uses a tensor product of a small number of $\mathcal{O}(\log n)$-qubit high dimensional signals \cite{PhysRevLett.87.167902}, whereas a recently-proposed optical protocol uses a tensor product of $\mathcal{O}(n)$ single-qubit signals, while maintaining the $\mathcal{O}(\log n)$ information leakage of the original protocol \cite{PhysRevA.89.062305}. We find a family of protocols which interpolate between the original and optical protocols while maintaining the $\mathcal{O}(\log n)$ information leakage, thus demonstrating a trade-off between the number of signals sent and the dimension of each signal.

There has been interest in experimental realization of the recently-proposed optical protocol using coherent states \cite{Xu:2015aa, PhysRevLett.116.240502}, but as the required number of laser pulses grows linearly with the input size $n$, eventual challenges for the long-time stability of experimental set-ups arise. We find a coherent state protocol which reduces the number of signals by a factor $1/2$ while also reducing the information leakage. Our reduction makes use of a simple modulation scheme in optical phase space, and we find that more complex modulation schemes are not advantageous. Using a similar technique, we improve a recently-proposed coherent state protocol for evaluating the Euclidean distance between two real unit vectors \cite{PhysRevA.95.032337} by reducing the number of signals by a factor $1/2$ and also reducing the information leakage.
\end{abstract}

\maketitle

%\tableofcontents

%%%%%%%%%%%%%%%%%%%%%%%%%%%%%%%%%%%%%%%%%%%%%%%%%%%%%
\section{\label{sec:level1} Introduction}
%%%%%%%%%%%%%%%%%%%%%%%%%%%%%%%%%%%%%%%%%%%%%%%%%%%%%

In this work we introduce several families of equality protocols in the simultaneous message passing model. In this model, two parties (Alice and Bob) receive inputs $x, y \in \{0,1\}^n$ respectively, conditioned on which they each send classical or quantum states to the referee, who performs a measurement to determine the output of some function $f(x,y)$. We consider the case in which $f$ is the equality function. We are interested in protocols which minimize the information leakage, that is, the amount of information the referee learns about the parties' inputs. Using classical states, the information leakage is lower bounded by $\Omega(\sqrt{n})$ \cite{1607.07516}. In contrast, there exist protocols using quantum states with information leakage $\mathcal{O}(\log n)$ \cite{PhysRevLett.87.167902, PhysRevA.89.062305}.

The original quantum fingerprinting protocol uses $\mathcal{O}(\log n)$-qubit highly entangled signals and a controlled-swap measurement \cite{PhysRevLett.87.167902}. A more recent and experimentally realizable ``optical'' protocol uses a tensor product of $\mathcal{O}(n)$ binary phase-modulated laser pulses, which can be represented as qubits, and a beamsplitter comparison measurement \cite{PhysRevA.89.062305}. In this work, we find a family of protocols which interpolate between these two, exhibiting a trade-off between the number of signals sent and the dimension of each signal. We show that this family of protocols has information leakage $\mathcal{O}(\log n)$.

The optical protocol of \cite{PhysRevA.89.062305} has been implemented using coherent states \cite{Xu:2015aa, PhysRevLett.116.240502}, but as the required number of laser pulses grows linearly with the input size $n$, large inputs become difficult to handle due to the limited long-time stability of experimental set-ups. We reduce the number of signals by a factor $1/2$, while also reducing the information leakage, by utilizing the imaginary component of the phase space representation of coherent states. We find several natural generalizations of this protocol which further reduce the number of signals, but find numerical evidence that the information leakage of these protocols is higher in both the ideal and experimental settings, even under an abstract, optimal measurement performed by the referee.

Using a similar technique, we also reduce the number of signals and information leakage of a recently-proposed coherent state protocol for evaluating the Euclidean distance between two real unit vectors \cite{PhysRevA.95.032337}, and find a similar protocol for complex unit vectors.

In addition, in Appendix~\ref{beamcomp} we prove a tangentially related result that a simple beamsplitter measurement similar to that used in \cite{andersson2006experimentally} achieves optimal unambiguous state comparison (USC) between any two coherent states of equal amplitude and opposite phase when the states are given with equal a priori probabilities. Optimal USC of two states given with equal a priori probabilities was first solved in \cite{BARNETT2003189} and generalized to arbitrary a priori probabilities in \cite{kleinmann2005generalization}. A method to realize the optimal USC of two single-photon states prepared with arbitrary a priori probabilities is proposed in \cite{qing2009linear}, but to our knowledge optimal USC has not yet been experimentally realized. The beamsplitter scheme has the advantage of being experimentally realizable, with the drawback of being sub-optimal for not-equal a priori probabilities.

%The comparison of squeezed vacuum states has been considered in \cite{PhysRevA.83.012313}.

The main text is organized as follows. In Section~\ref{int} we interpolate between the original and existing optical equality protocols. In Section~\ref{optical} we introduce our coherent state protocols which improve on the existing optical equality and Euclidean distance protocols, and find numerical evidence that some natural generalizations to coherent state equality protocols using fewer signals have higher information leakage. In Section~\ref{infleaksection} we derive the information leakage bounds we have used for our protocols.
% In Appendix~\ref{beamcomp} we describe a beamsplitter measurement to perform optimal unambiguous state comparison between two coherent states.

%%%%%%%%%%%%%%%%%%%%%%%%%%%%%%%%%%%%%%%%%%%%%%%%%%%%%
\section{Interpolation}\label{int}
%%%%%%%%%%%%%%%%%%%%%%%%%%%%%%%%%%%%%%%%%%%%%%%%%%%%%

The original equality protocol of \cite{PhysRevLett.87.167902} uses a small number of  $\mathcal{O}(\log n)$-qubit signals, whereas an existing optical equality protocol of \cite{PhysRevA.89.062305} uses a tensor product of $\mathcal{O}(n)$ binary phase-modulated coherent state signals, which can be represented as single qubits (a notion formalized in Section~\ref{adaptopt}). In this section we interpolate between the original and optical protocols, thus demonstrating a trade-off between the number of signals sent and the dimension of each signal. In Sections~\ref{adaptopt} and~\ref{adaptor} we introduce slight adaptations to the existing protocols which are more natural candidates for the interpolation, and in Section~\ref{intc} we interpolate between these adaptations.

Before proceeding, we outline a general protocol framework which we will use for all equality protocols that we consider. First, Alice and Bob receive inputs $x, y \in\{0,1\}^n$ respectively, conditioned on which they send pure states $\kett{\psi_x}, \kett{\psi_{y}}$ to the referee which are sufficiently distinguishable when $x \neq y$. The referee then performs a comparison measurement on $\kett{\psi_{xy}}:=\kett{\psi_x}\kett{\psi_{y}}$ and outputs either Equal or NotEqual. We define the error probability of the protocol as the worst case error probability over all $x,y \in \{0,1\}^n$. In the ideal setting, the error probability of every protocol is one-sided: if the inputs are equal the referee will always output Equal. In every protocol, the states $\kett{\psi_x}$ are product vectors. We refer to individual tensor factors of $\kett{\psi_x}$ as \textit{signals}, and to the entire object $\kett{\psi_x}$ as a \textit{state}. For many protocols that we consider, the states will contain multiple copies of identical signals.

To make the states sufficiently distinguishable to the referee when $x \neq y$, Alice and Bob map their inputs $x, y$ to codewords $\tE(x),\tE(y)\in \{0,1\}^m$ of an error-correcting code $\tE$ characterized by some minimum distance. They then encode these codewords into states $\kett{\psi_x}, \kett{\psi_{y}}$ whose overlap is a decreasing function of the distance between codewords. The code $\tE$ is chosen to have constant minimum distance $m \delta$ and constant rate, which we will see ensures that the states used in each protocol are sufficiently distinguishable to the referee and give rise to information leakage $\mathcal{O}(\log n)$.

\subsection{Adaptation of existing optical equality protocol}\label{adaptopt}
%%%%%%%%%%%%%%%%%%%%%%%%%%%%%%%%%%%%%%%%%%%%%%%%%%%%%
Now we review the existing optical equality protocol of \cite{PhysRevA.89.062305} (in the ideal setting) and propose a slight adaptation which is a more natural candidate for the interpolation.
In the existing optical protocol, the $j$-th signal is one of two coherent states depending on the $j$-th codeletter of the codeword $\tE(x)\in\{0,1\}^m$ \cite{PhysRevA.89.062305}
\begin{align}\label{state:original}
\ket{\alpha_x^{}}_{\text{EQ},1}=\bigotimes_{j=1}^m \ket{(-1)^{\tE(x)_j}\frac{\alpha}{\sqrt{m}}}_j.
\end{align}
%Note that the standard basis vectors $\ket{j}$ appearing in \eqref{state:original} are unnecessary (and unused in practice), but we include them here 
For each index $j$, the referee interferes the $j-$th pair of signals received from Alice and Bob in a beamsplitter, and measures the dark port with a single photon threshold detector, obtaining one of two outcomes: ``dark port detection'' or ``no dark port detection''. The referee outputs NotEqual if at least one outcome ``dark port detection'' occurs. On input $\ket{\beta_a}\ket{\beta_b}$, outcome ``no dark port detection'' occurs with probability
\begin{align}\label{beamerr}
\abs{\bk{\beta_a}{\beta_b}}=e^{-\frac{1}{2}\abs{\beta_a-\beta_b}^2}.
\end{align}
It follows that the error probability given different inputs $x\neq y$ is equal to $\abss{\bkk{\alpha_x}{\alpha_{y}}}$, and the error probability given equal inputs is zero. The worst case error probability occurs when the codewords differ by minimum distance $m \delta$ bits, and is equal to $\exp[-2 \abs{\alpha}^2 \delta]$, which is brought to within any $\epsilon>0$ through appropriate choice of $\alpha$.

In the existing optical protocol, the set of two possible coherent states for each signal span a two-dimensional space, and thus can be written in a basis as (i.e. they are isometrically equivalent to) two qubits $\kett{q_0^\epsilon},\kett{q_1^\epsilon}$ with inner product determined by $\epsilon$. The beamsplitter measurement can also be converted to this basis (see Appendix~\ref{beamcomp}). By the invariance of entropy under isometries, the information leakage (defined in Section~\ref{infleaksection}) is equal for protocols using states that are equal up to change of basis.

In contrast to the existing optical protocol, the original protocol of \cite{PhysRevLett.87.167902} attains the desired error probability $\epsilon$ by sending multiple copies of signals which are fixed independent of $\epsilon$. To facilitate our interpolation between these two protocols, we adapt the optical protocol to more closely resemble the original by fixing qubits $\kett{q_0'}, \kett{q_1'}$ independent of $\epsilon$, and sending identical copies of each qubit signal until $\epsilon$ is attained (we refer to each individual copy as a signal). In the coherent state basis, this corresponds to fixing the amplitude of each signal independent of $\epsilon$ (any two qubits can be written in a basis as coherent states). Define the \textit{repetition number} $r$ as the number of copies sent to attain $\epsilon$. The referee uses the qubit-basis beamsplitter measurement described in Appendix~\ref{beamcomp} on each signal, and outputs NotEqual if any outcome ``dark port detection'' occurs.

Remarkably, as the probability \eqref{beamerr} of ``no dark port detection'' is given by the overlap between the input pair of signals, it follows that if $\epsilon$ is attained with equality in both the existing optical protocol and our adapted protocol, then the states used in each protocol are equal up to a change of basis. Specifically, the set of signals $\{\kett{\frac{\alpha}{\sqrt{m}}},\kett{\frac{-\alpha}{\sqrt{m}}}\}$ used in the existing optical protocol are equal up to a change of basis to the set of unit vectors $\{\kett{q_0'}^{\otimes r}, \kett{q_1'}^{\otimes r}\}$ containing the $r$ copies of each signal used in our adapted protocol. Indeed, the worst case error probability is given by $\epsilon=\abss{\bkk{q'_0}{q'_1}}^{m \delta r}=\abss{\bkk{\frac{-\alpha}{\sqrt{m}}}{\frac{\alpha}{\sqrt{m}}}}^{m \delta}$, so $\abss{\bkk{q'_0}{q'_1}}^{r}=\abss{\bkk{\frac{-\alpha}{\sqrt{m}}}{\frac{\alpha}{\sqrt{m}}}}^{}$, which completes the proof by Property 3 of \cite{doi:10.1142/S0219749904000031} that two sets of unit vectors $\{\ket{v_a}\in\mathcal{H}_v\}_{a\in Z}, \{\ket{w_a}\in \mathcal{H}_w \}_{a\in Z}$ are equal up to change of basis if and only if there exist real numbers $\theta_a, a\in Z$ such that $\bk{v_a}{v_b}=e^{i(\theta_a-\theta_b)} \bk{w_a}{w_b}$ for all $a,b\in Z$.
Note that the choice of coherent state amplitude $\frac{\alpha}{\sqrt{m}}$ for the optical protocol thus corresponds to the choice $\frac{\alpha}{\sqrt{r m}}$ for our adapted optical protocol in the coherent state basis.

%%%%%%%%%%%%%%%%%%%%%%%%%%%%%%%%%%%%%%%%%%%%%%%%%%%%%
\subsection{Adaptation of original equality protocol}\label{adaptor}
%%%%%%%%%%%%%%%%%%%%%%%%%%%%%%%%%%%%%%%%%%%%%%%%%%%%%

Here we describe the original quantum fingerprinting protocol of \cite{PhysRevLett.87.167902}, and propose a slight adaptation which is a more natural candidate for the interpolation. Similarly to the adapted optical protocol, in the original protocol Alice and Bob send identical copies of the signals
\begin{align}\label{orstates}
\ispa{x}{}{m}&=\frac{1}{\sqrt{m}}\sum_{i=1}^{m} \ket{i}\ket{\tE(x)_i}
%&=\frac{1}{\sqrt{m}} \sum_{i=1}^m \ket{i} \ket{q^{(m)}_{\tE(x)_i}}
\end{align}
until the desired error probability $\epsilon$ is attained. On each pair of signals $\ispa{xy}{}{m}:=\ispa{x}{}{m}\ispa{y}{}{m}$ sent by Alice and Bob, the referee performs a controlled-swap measurement (effectively a projective measurement onto the symmetric and anti-symmetric subspaces), which returns outcome ``anti-symmetric'' with probability $\Vert \frac{1}{2} (\mathds{1}-\mathrm{W}) \ispa{xy}{}{m}\Vert^2$, where $\mathrm{W}$ is the operator which swaps the state of Alice and Bob's systems. The referee outputs NotEqual if any outcome ``anti-symmetric'' occurs. The worst case error probability occurs when the codewords differ by minimum distance $m \delta$ bits, and is given by  $(1-\delta(1-\frac{\delta}{2}))^r$ for repetition number $r$ \cite{PhysRevLett.87.167902}.

In our adapted original protocol, the signals used are the same but the referee instead performs a direct sum of the controlled-swap measurement with the qubit-basis beamsplitter measurement. We use this measurement for the interpolation because on one end, in the adapted optical protocol, it converges to the beamsplitter measurement (see Section~\ref{intc}); and on the other end, in the adapted original protocol, it closely resembles (and slightly improves on) the controlled-swap measurement.

Operationally, the measurement in the adapted original protocol proceeds as follows. First, the referee performs a non-destructive measurement on $\ispa{xy}{}{m}$ with two measurement operators, the first (second) of which projects onto the subspace containing the first (second) term of the decomposition
\begin{align}\label{decomp}
\ispa{xy}{}{m}=&\frac{1}{m}\sum_{i=1}^m \ket{i}\ket{\tE(x)_i}\ket{i}\ket{\tE(y)_i}\nonumber\\
&+\frac{1}{m}\sum_{\substack{l,h=1 \\ l\neq h}}^m\ket{l}\ket{\tE(x)_l}\ket{h}\ket{\tE(y)_h}.
\end{align}
%In the original protocol, 
%\begin{align}
%\left[1-\delta\left(1-\frac{\delta}{2}\right)\right]^r.
%\end{align}
If the non-destructive measurement projects onto the first subspace, the referee performs a measurement which applies the identity matrix to the first and third registers, and to the two-qubit space contained in the second and fourth registers it applies the measurement operators $M_{\text{d}}$ and $M_{\text{nd}}$, implicitly defined in Appendix~\ref{beamcomp}, which correspond to the measurement outcomes ``dark port detection'' and ``no dark port detection'', respectively. If the measurement projects onto the second subspace, the referee performs the controlled-swap measurement on the entire state. The referee outputs NotEqual if any outcome ``dark port detection'' or ``anti-symmetric'' occur. In Appendix~\ref{interperr} we show that this protocol has worst case error probability
\begin{align}\label{orpsame}
{\Pr}_m^\tI(\text{Err})=\left[1-\delta\left(1-\frac{\delta}{2}+\frac{1}{2m}\right)\right]^r,
\end{align}
a minor improvement over the original protocol.

In Appendix~\ref{beamcomp} we show that the qubit-basis beamsplitter measurement on the two-qubit space contained in the second and fourth registers can be further decomposed as a direct sum of the controlled-swap measurement with an unambiguous state discrimination measurement. Thus, the full adapted measurement on the larger Hilbert space can also be decomposed in this way.

%%%%%%%%%%%%%%%%%%%%%%%%%%%%%%%%%%%%%%%%%%%%%%%%%%%%%
\subsection{Interpolation between adapted protocols}\label{intc}
%%%%%%%%%%%%%%%%%%%%%%%%%%%%%%%%%%%%%%%%%%%%%%%%%%%%%
Now we find a family of protocols which interpolates between the adapted protocols described in Sections~\ref{adaptopt} and ~\ref{adaptor}, thus demonstrating a trade-off between the number of signals sent and the dimension of each signal. In the interpolation protocol with block size $k$, blocks of $k$ bits of $\tE(x)$ are encoded into each signal:
\begin{align}\label{instates}
\ispa{x}{}{k}=\bigotimes_{j=1}^{\cl{m/k}} \left[\frac{1}{\sqrt{k}}\sum_{i\in \tI[j,k]} \ket{i}\ket{q^{(k)}_{\tE(x)_i}}\right] \in (\mathds{C}^{k}\otimes \mathds{C}^2)^{\otimes\cl{\frac{m}{k}}},
\end{align}
where $\kett{q^{(k)}_0},\kett{q^{(k)}_1}$ are qubits. The set $\tI[j,k]$ indexes the $j$-th block of $k$ bits, i.e. it is the set of integers in the range $[(j-1)k+1,jk]$. If $k$ does not divide $m$ the remaining qubits in the final signal are set to $\kett{q^{(k)}_0}$. As before, Alice and Bob send identical copies of $\ispa{x}{}{k}$ until the desired error probability $\epsilon$ is attained. Note that Eq. \eqref{instates} converges to Eq. \eqref{orstates} for $k=m$, and for $k=1$ it converges to
\begin{align}
\ispa{x}{}{1}=\bigotimes_{j=1}^m \ket{j}\ket{q^{(1)}_{\tE(x)_j}},
\end{align}
which reproduces Eq. \eqref{state:original} written in the qubit basis and appended with basis vectors $\ket{j}$.
%As evidenced by \eqref{instates}, for each $k=1,\dots, m$ our interpolation uses $r \cl{m/k}$ signals, each of dimension $2k$. This demonstrates a trade-off between the number of signals sent and the dimension of each signal.

% interpolation demonstrating a trade-off between the adapted original protocol, which uses $r$ signals of dimension $2m$, and the adapted optical protocol, which uses $r m$ signals of dimension two.

For each index $j=1,\dots,\cl{m/k}$, the referee measures the $j$-th pair of signals received from Alice and Bob as follows. As in the adapted original protocol, they perform a direct sum of the qubit-basis beamsplitter measurement and the controlled-swap measurement on the first and second terms of the decomposition
\begin{align}\label{instatesdecomp}
&\frac{1}{k}\sum_{i\in \tI[j,k]} \left[\ket{i}\ket{q^{(k)}_{\tE(x)_i}}\ket{i}\ket{q^{(k)}_{\tE(y)_i}}\right]\nonumber\\
&+\frac{1}{k}\sum_{\substack{l,h \in \tI[j,k] \\ l\neq h}}\left[\ket{l}\ket{q^{(k)}_{\tE(x)_l}}\ket{h}\ket{q^{(k)}_{\tE(y)_h}}\right],
\end{align}
and decide NotEqual on any outcome ``dark port detection'' or ``anti-symmetric''. For $k=m$ this measurement converges to the measurement used in the adapted original protocol, and for $k=1$ the second term of the decomposition \eqref{instatesdecomp} disappears and this measurement converges to the qubit-basis beamsplitter measurement. The worst case error probability of this measurement is derived in Appendix~\ref{interperr}.

As noted in the previous section, this measurement can equivalently be decomposed as a direct sum of the controlled-swap measurement with an unambiguous state discrimination measurement.

As shown in Eq. \eqref{instates}, each signal contains $k$ qubit states, each chosen from the set $\{\kett{q^{(k)}_0},\kett{q^{(k)}_1}\}$. 
%We will choose these qubits to satisfy $\abss{\bkk{q^{(k)}_0}{q^{(k)}_1}}=1-k/m$ as one example which converges to the adapted orginal and optical protocols for $k=m$ and $k=1$ respectively. In Section~\ref{intinf} we show that the information leakage under this choice is $\mathcal{O}(r \log n)$, and in Appendix~\ref{interperr} we show that any fixed error can be attained with constant repetition number $r$.
% Furthermore, for fixed error $\epsilon$, this choice gives rise to fixed repetition number $r$, which allows us in Section~\ref{intinf} to bound the information leakage as $\mathcal{O}(\log n)$ for every block-size $k$.
%In summary, this family of protocols interpolates between the adapted original and optical protocols while maintaining information leakage $\mathcal{O}(\log n)$.
%For a given block size $k$, our interpolation uses $r \cl{m/k}$ signals, each of dimension $2k$, thus demonstrating a linear relationship between the number of signals sent and the dimension of each signal.
We will choose these qubits to satisfy ${\bkk{q^{(k)}_0}{q^{(k)}_1}}=1-k/m$, which converges to the adapted orginal and optical protocols for $k=m$ and $k=1$ respectively, and which has information leakage $\mathcal{O}(\log n)$ (see Section~\ref{intinf}).
% Furthermore, for fixed error $\epsilon$, this choice gives rise to fixed repetition number $r$, which allows us in Section~\ref{intinf} to bound the information leakage as $\mathcal{O}(\log n)$ for every block-size $k$.
%In summary, this family of protocols interpolates between the adapted original and optical protocols while maintaining information leakage $\mathcal{O}(\log n)$.
For a given block size $k$ and repetition number $r$, our interpolation uses $r \cl{m/k}$ signals, each of dimension $2k$, exhibiting a trade-off between the number of signals sent and the dimension of each signal.

%%%%%%%%%%%%%%%%%%%%%%%%%%%%%%%%%%%%%%%%%%%%%%%%%
\section{Optical Protocols}\label{optical}
%%%%%%%%%%%%%%%%%%%%%%%%%%%%%%%%%%%%%%%%%%%%%%%%%

In this section we consider several families of optical coherent state simultaneous message passing model protocols which reduce the number of signals below that of the existing protocols. In Section~\ref{imp_opt} we introduce optical protocols for equality and Euclidean distance which reduce the number of signals by a factor $1/2$ and reduce the information leakage below that of the existing optical protocols of \cite{PhysRevA.89.062305} and \cite{PhysRevA.95.032337}. In Section~\ref{opt_eq_families} we introduce two families of optical equality protocols which further reduce the number of signals, but find numerical evidence that they increase the information leakage in both the ideal and experimental settings, even under an abstract, optimal measurement performed by the referee.

\subsection{Improved optical protocols for equality and Euclidean distance}\label{imp_opt}

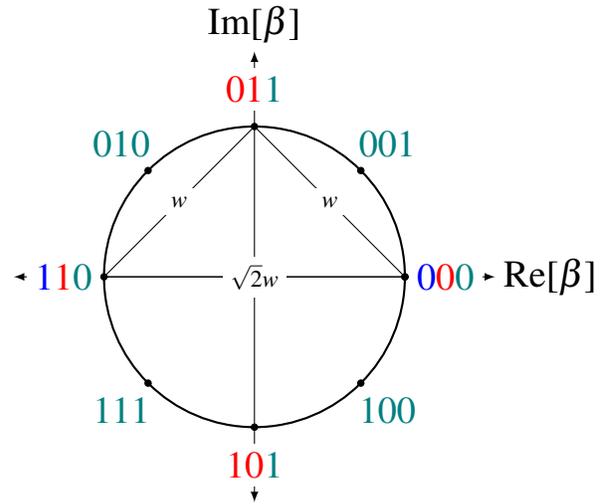
\begin{figure}[H]
\begin{tikzpicture}[scale=2,cap=round,>=latex]
        % draw the coordinates
        \draw[<->] (-1.6cm,0cm) -- (1.6cm,0cm) node[right,fill=white] {{\Large Re[$\beta$]}};
        \draw[<->] (0cm,-1.5cm) -- (0cm,1.5cm) node[above,fill=white] {\Large Im[$\beta$]};
\draw[-](-1cm,0cm) -- (0cm,1cm);
\draw[-](1cm,0cm) -- (0cm,1cm);
\draw[-](-1cm,0cm) -- (1cm,0cm);
\draw (-.5cm,.5cm) node[fill=white]{$w$};
\draw (.5cm,.5cm) node[fill=white]{$w$};
\draw (0cm,0cm) node[fill=white]{$\sqrt{2}w$};
        % draw the unit circle
        \draw[thick] (0cm,0cm) circle(1cm);

        \foreach \x in {0,45,...,360} {
                % lines from center to point
%                \draw[gray] (0cm,0cm) -- (\x:1cm);
                % dots at each point
                \filldraw[black] (\x:1cm) circle(0.6pt);
                % draw each angle in degrees
%                \draw (\x:0.6cm) node[fill=white] {$\x^\circ$};
        }

        \draw (-1.27cm,0cm) node[fill=white] {{\Large \blue{1}\red{1}\green{0}}}
              (1.27cm,0cm)  node[fill=white] {{\Large \blue{0}\red{0}\green{0}}}
              (0cm,-1.25cm) node[fill=white] {{\Large \red{1}\red{0}\green{1}}}
              (0cm,1.25cm)  node[fill=white] {{\Large \red{0}\red{1}\green{1}}}
	(.883883cm,.883883cm) node[fill=none]{{\Large \green{0}\green{0}\green{1}}}
(.883883cm,-.883883cm) node[fill=none]{{\Large \green{1}\green{0}\green{0}}}
(-.883883cm,.883883cm) node[fill=none]{{\Large \green{0}\green{1}\green{0}}}
(-.883883cm,-.883883cm) node[fill=none]{{\Large \green{1}\green{1}\green{1}}};
    \end{tikzpicture}
\caption{Gray coding of $k-$bit blocks into a ring of coherent states in phase space for $k=1$ (blue), $k=2$ (blue and red combined), and $k=3$ (blue, red and teal combined). Here, $\beta$ is the coherent state amplitude of each signal. For $k=1$, $\beta_1=\alpha/\sqrt{m}$; for $k=2$, $\beta_2=\alpha/\sqrt{m/2}$; and in general, $\beta_k=\sqrt{\mu_k/(m/k)}$, where $\mu_k$ is the total mean photon number of the total state. \label{psk_figure}}
\end{figure}

In our improved optical equality protocol, two bits of $\tE(x)$ are encoded into each signal by utilizing the imaginary component of the phase space representation of coherent states, which reduces the number of signals by a factor $1/2$. Codeletters $01/10$ are encoded into phases $\pm i$, and codeletters $00/11$ are encoded into phases $\pm1$, as shown in Figure~\ref{psk_figure}. Explicitly, the parties send the states
\begin{align}\label{state:phase}
\ket{\alpha_x}_{\text{EQ},2}=\bigotimes_{\substack{j=1\\\text{odd}}}^{m} \ket{(-1)^{\tE(x)_j}\cdot (i)^{\tE(x)_j\oplus \tE(x)_{j+1}}\frac{\alpha}{\sqrt{m/2}}}_j.
\end{align}
The referee uses the same beamsplitter measurement as in the existing optical protocol: she interferes pairs of signals in a beamsplitter, measures the dark port with a single photon threshold detector, and decides NotEqual if at least one outcome ``dark port detection'' occurs. As before, the desired error probability $\epsilon$ is attained through appropriate choice of $\alpha$.

The above states have the same total mean photon number $\abs{\alpha}^2$, and give rise to the same error probability, as the existing optical protocol. To show the second statement, recall that the probability of ``no dark port detection'' 
depends only on the squared distance between the amplitudes of the incoming pair of coherent state signals \eqref{beamerr}. For pairs of codeletters $(\tE(x)_j, \tE(x)_{j+1})$ and $(\tE(y)_j, \tE(y)_{j+1})$ which differ by one bit this quantity is given by $w^2={\abs{\alpha}^2}/{m}$ as in the existing optical protocol (see Figure~\ref{psk_figure}), and for pairs which differ by two bits this quantity is $2 w^2$, which gives rise to the same probability of ``no dark port detection'' as the existing optical protocol (see Appendix~\ref{optenc}). By the information leakage bound ${\sim}~{\mathcal{O}(\abs{\alpha}^2 \log m_k)}$ (where $m_k$ is the number of signals) derived in Appendix~\ref{optinf}, our improved protocol has lower information leakage than the existing protocol. It can be shown that this statement also holds under the stronger bound derived in Section~\ref{impoptinf} for $\abss{\alpha}^2\ll m_k$ using standard approximation techniques. Below we will refer to this protocol, including its use of the beamsplitter measurement, as the two-bit protocol.

We improve the existing optical Euclidean distance protocol of \cite{PhysRevA.95.032337} in similar fashion. In the existing protocol, Alice and Bob receive real unit vectors $u,v \in \mathds{R}^s$ respectively and prepare the states
\begin{align}\label{state:EDoriginal}
\ket{\alpha_u}_{\text{ED},1}:=\bigotimes_{j=1}^s \vert u_j \alpha \rangle_j.
\end{align}
The referee interferes each pair of signals received from Alice and Bob in a beamsplitter and measures both output ports with single photon threshold detectors. The quantity $\Vert u-v \Vert^2$ is a function of $\abs{\alpha}^2$ and the expected difference between the number of detections observed in the two output ports, so using Chernoff bounds the referee can estimate $\Vert u-v \Vert^2$ to within an additive constant $\epsilon$ with probability at least $1-\delta$ by repeating the protocol $\mathcal{O}(\log(1/\delta)/\epsilon^2)$ times \cite{PhysRevA.95.032337}.

As in our improved equality protocol, our improved Euclidean distance protocol utilizes the imaginary component of the phase space representation of coherent states to reduce the number of signals by a factor $1/2$. Alice and Bob prepare the states
\begin{align}\label{state:EDphase}
\ket{\alpha_u}_{\text{ED},2}:=\bigotimes_{\substack{j=1\\ \text{odd}}}^{s} \vert (u_j+iu_{j+1}) \alpha \rangle_j,
\end{align}
and the referee uses the same measurement as before.  These states have the same total mean photon number $\abs{\alpha}^2$, and it can be shown using nearly identical analysis to \cite{PhysRevA.95.032337} that the measurement outcome statistics are also the same. Thus, as before, our protocol has lower information leakage than the existing protocol under the information leakage bound of Appendix~\ref{optinf}. Alternatively, one can adapt the existing protocol to evaluate the Euclidean distance between two complex unit vectors $u, v\in \mathds{C}^s$ using the same measurement and the states \eqref{state:EDoriginal}.

\subsection{Families of optical equality protocols}\label{opt_eq_families}
In this section we introduce two families of optical coherent state equality protocols which further reduce the number of signals. In Section~\ref{optideal} we find numerical evidence that these protocols have higher information leakage than the two-bit protocol in the ideal setting, even under the optimal measurement. In Section~\ref{optexp} we find numerical evidence of the same behaviour under realistic experimental imperfections.

% For our optical protocols with block-size $k$, blocks of $k$ bits of $\tE(x)$ are encoded into one of $2^k$ coherent state signals forming a ring or lattice in phase space using a Gray code (see Figure~\ref{psk_figure}).
%% using a Gray code, which satisfies the property that nearest neighbour nodes differ by one bit. The ring Gray code is shown in Figure~\ref{psk_figure}.
%In Appendix~\ref{optenc} we prove that this is the optimal such encoding for all $k=2,3,4$, and that it outperforms an analogous encoding of $q-$ary codewords.
%%for all $k=2,3,4$ the Gray code is the optimal such encoding in that it minimizes the worst case error probability of the beamsplitter measurement. We also prove that the Gray coding outperforms an analogous encoding of $d-$ary codewords into a ring or lattice.
%
%In Appendix~\ref{opmeas} we derive the optimal measurement for any one-sided error (i.e. zero error for equal inputs) simultaneous message passing model equality protocol, and find that for the optical protocols the worst case error probability of the optimal measurement is lower bounded by the square of the error probability of the beamsplitter measurement. In the following sections we use this bound to find numerical evidence that even under the optimal measurement, the higher-bit protocols transmit more information than the two-bit protocol in the ideal setting. We also find numerical evidence of this behaviour in the experimental setting under the beamsplitter measurement.

%%%%%%%%%%%%%%%%%%%%%%%%%%%%%%%%%%%%%%%%%%%%%%%%%%%%%
\subsubsection{Ideal setting}\label{optideal}
%%%%%%%%%%%%%%%%%%%%%%%%%%%%%%%%%%%%%%%%%%%%%%%%%%%%%

We first describe our families of optical equality protocols in the ideal setting. In the ring (lattice) protocol with block size $k$, blocks of $k$ bits of $\tE(x)$ are encoded into one of $2^k$ coherent state signals arranged in a ring (lattice) in phase space using a Gray code, as shown in Figure~\ref{psk_figure}. The size of the ring (lattice) is determined by the desired error probability $\epsilon$, and is held constant for all signals.

The ring (lattice) Gray code is a mapping from $k$-bit strings to a ring (lattice) of coherent state signals which satisfies the property that all nearest neighbour signals differ in exactly one bit \cite{gray1953pulse, DeOliveira1992, campopiano1962coherent}. Note that in the ring encoding, each coherent state signal has two nearest neighbours, while in the lattice encoding a given coherent state signal can have as many as four nearest neighbours. We have chosen this code so that a greater number of bit differences between two $k$-bit blocks of $\tE(x)$ and $\tE(y)$ corresponds to greater distinguishability between the two coherent state signals. In Appendix~\ref{optenc} we prove that this is the optimal encoding of $k$-bit blocks of binary codewords for all $k=2,3,4$, and that it outperforms an analogous encoding of $q-$ary codewords.

We consider two different measurements performed by the referee. The beamsplitter measurement proceeds identically to that of Section~\ref{adaptopt}: the referee interferes pairs of signals in a beamsplitter and measures the dark port with a single photon threshold detector. She decides NotEqual if at least one outcome ``dark port detection'' occurs. Recall that the error probability under the beamsplitter measurement is one-sided, i.e. there is zero error for equal inputs. We also consider the optimal one-sided error measurement, which is described in Appendix~\ref{opmeas}, and is shown to have error probability lower bounded by the square of the error probability of the beamsplitter measurement.

%Now we describe Alice and Bob's signal states. In the ring (lattice) protocol with block size $k$, blocks of $k$ bits of $\tE(x)$ are encoded into one of $2^k$ coherent state signals arranged in a ring (lattice) in phase space using a Gray code, as shown in Figure~\ref{psk_figure}. The size of the ring (lattice) is determined by the referee's measurement and the desired worst case error probability $\epsilon$, and is held constant for all signals.

\begin{figure}
\includegraphics[scale=0.5]{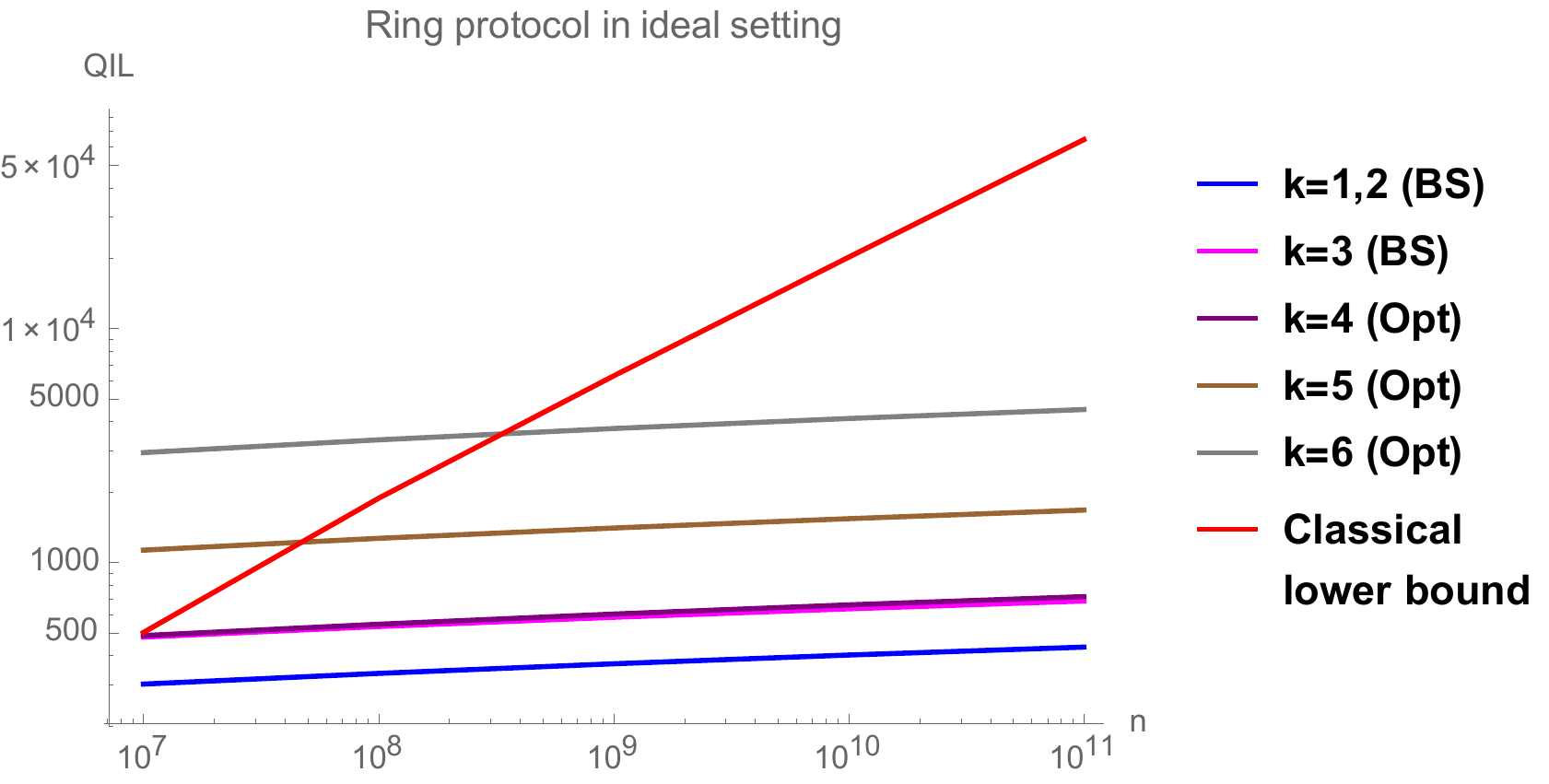}
\caption{Quantum information leakage (QIL), measured in bits, as a function of the input size $n$ for our ring protocols to attain error probability $\epsilon=0.01$ in the ideal setting. \label{QF_ideal}}
\end{figure}

In Figure~\ref{QF_ideal} we plot the information leakage of the ring encoding under the bound of Section~\ref{impoptinf}, compared to the classical information leakage lower bound of \cite{1607.07516}. We have optimized over $\delta$ and assumed the code $\tE$ saturates the Gilbert-Varshamov bound \cite{BLTJ:BLTJ1393, varshamov, van2013introduction}
\begin{align}\label{bingv}
\frac{n}{m} = 1-\text{h}(\delta).
\end{align}
For $k=1,2,3$ we plot the information leakage under the beamsplitter measurement, and for $k=4,5,6$ we lower bound the information leakage under the optimal measurement using the quadratic bound on the error probability derived in Appendix~\ref{opmeas}. We see that the one-bit and two-bit protocols have the lowest information leakage (they are numerically indistinguishable in this parameter regime). We have observed the same result for $\epsilon$ in the range $[10^{-5},10^{-2}]$ in both the ring and lattice protocols.

Before continuing on to consider experimental imperfections, we argue that for fixed block size $k$, both the ring and lattice protocols have asymptotic information leakage $\mathcal{O}(\log n)$ in the ideal setting. In Appendix~\ref{optinf} we show that any simultaneous message passing model protocol which uses $m_k$ coherent state signals and fixed maximum total mean photon number $\mu_{\mathrm{max},k}$ has information leakage ${\sim}~{\mathcal{O}(\mu_{\mathrm{max},k} \log m_k)}$. Note that $m_k=m/k=\mathcal{O}(n)$ by the fact that $\tE$ is a constant rate code. Furthermore, in Appendix~\ref{optenc} we show that any fixed error probability is attained with $\mu_{\mathrm{max},k}$ constant in $n$. Together, these results imply that the information leakage is $\mathcal{O}(\log n)$.
%First we upper bound the error probability given different inputs $x\neq y \in \{0,1\}^n$. For any such inputs, note that a single pair of coherent state signals sent from Alice and Bob can differ in at most $k$ bits, so at least $m \delta/k$ pairs of signals will differ. By the geometry of the ring (lattice) the overlap between any two different signals is exponentially small in the square of the (largest) signal amplitude $\abs{\beta}^2$. Thus, the worst case error probability is exponentially small in $\abs{\beta}^2 m \delta/k=\delta \mumax$, and is brought to within any $\epsilon>0$ with fixed $\mumax$. See Appendix~\ref{optenc} for a detailed calculation.

%%%%%%%%%%%%%%%%%%%%%%%%%%%%%%%%%%%%%%%%%%%%%%%%%%%%%
\subsubsection{Experimental imperfections}\label{optexp}
%%%%%%%%%%%%%%%%%%%%%%%%%%%%%%%%%%%%%%%%%%%%%%%%%%%%%
In the experimental setting, for larger block sizes the states have fewer signals, so dark counts have less effect on the error probability. In this section we find numerical evidence that despite this effect, the two-bit protocol still outperforms the ring protocols with block size $k>2$ in the experimental setting, and speculate that the same result holds for the lattice protocols.

The experimental ring protocol uses the same states and beamsplitter measurement as in the ideal setting. However, to account for transmittivity $\eta$ the initial total mean photon number $\mu_k$ is rescaled to $\mu_k/\eta$, and to account for dark count probability $p_{\text{dark}}$ per signal the referee uses a different criteria to decide Equal or NotEqual. The referee decides NotEqual if the number of outcomes ``dark port detection'' exceeds a threshold value $\dthresh$ which is chosen to minimize the worst case error probability over all inputs $x,y \in \{0,1\}^n$ (which is no longer one-sided when $p_{\text{dark}}>0$). This technique is based on that introduced in \cite{Xu:2015aa} for experimental implementation of the existing optical protocol.

%This criteria converges to that of the ideal setting $(\dthresh=1)$, which has one sided worst case error. In contrast, when $p_{\text{dark}}>0$ the error is no longer one-sided.

Now we determine the optimal threshold value $\dthresh$. Define a random variable $D_{E,k}$ for the number of outcomes ``dark port detection'' given equal inputs. Define $D_{D,k}$ identically, but for different inputs which have the lowest expected number of outcomes ``dark port detection''. For a given threshold value $\dthresh'$, the worst case error probability is then given by $\max\{\Pr(D_{E,k}\geq \dthresh'),\Pr(D_{D,k}< \dthresh')\}$. As the first (second) element is monotonically decreasing (increasing) with $\dthresh'$, it follows that the optimal threshold value $\dthresh$ satsifies
\begin{align}\label{dthresh}
\Pr(D_{E,k}\geq \dthresh)=\Pr(D_{D,k}< \dthresh),
\end{align}
which is also the worst case error probability of the protocol under this choice. Note that Eq. \eqref{dthresh} may not attain exact equality due to the fact the threshold value must be an integer, so both probabilities are step functions.

In our considered parameter regime, it can be shown that the number of clicks are well-approximated by binomial distributions $D_{D,k} \sim \text{Bin}(m/k, p_{D,k})$ and $D_{E,k}\sim \text{Bin}(m/k, p_{E,k})$, where
\begin{align}\label{pdk}
p_{D,k}&=(1-(k \delta-\fl{k \delta})) \left(1-e^{-\abs{\beta_k}^2\left[1-\cos\left(\frac{2\pi}{2^k}\fl{k \delta}\right)\right]}\right)\nonumber\\
&+(k \delta-\fl{k \delta})\left(1-e^{-\abs{\beta_k}^2\left[1-\cos\left(\frac{2\pi}{2^k}\left(\fl{k \delta}+1\right)\right)\right]}\right)+p_{\text{dark}}\nonumber\\
p_{E,k}&=p_{\text{dark}}
\end{align}
%\begin{align}\label{pdk}
%p_{D,k}&=1-(1-(k \delta-\fl{k \delta}))e^{-
%%\frac{\mu_k}{m/k}
%\beta_k \left[1-\cos\left(\frac{2\pi}{2^k}\fl{k \delta}\right)\right]}\nonumber\\
%&\hspace{.4in}-(k \delta-\fl{k \delta})e^{-
%%\frac{\mu_k}{m/k}
%\beta_k\left[1-\cos\left(\frac{2\pi}{2^k}\left(\fl{k \delta}+1\right)\right)\right]}+p_{\text{dark}}\nonumber\\
%p_{E,k}&=p_{\text{dark}}
%\end{align}
for all $k=1,\dots, 6$ under the Gray code (see Appendix~\ref{optenc}), where $\beta_k=\sqrt{\mu_k/(m/k)}$ is the amplitude of each coherent state signal. We have used this approximation to calculate $\dthresh$ and the corresponding worst case error probability \eqref{dthresh}.

\begin{figure}
\includegraphics[scale=0.5]{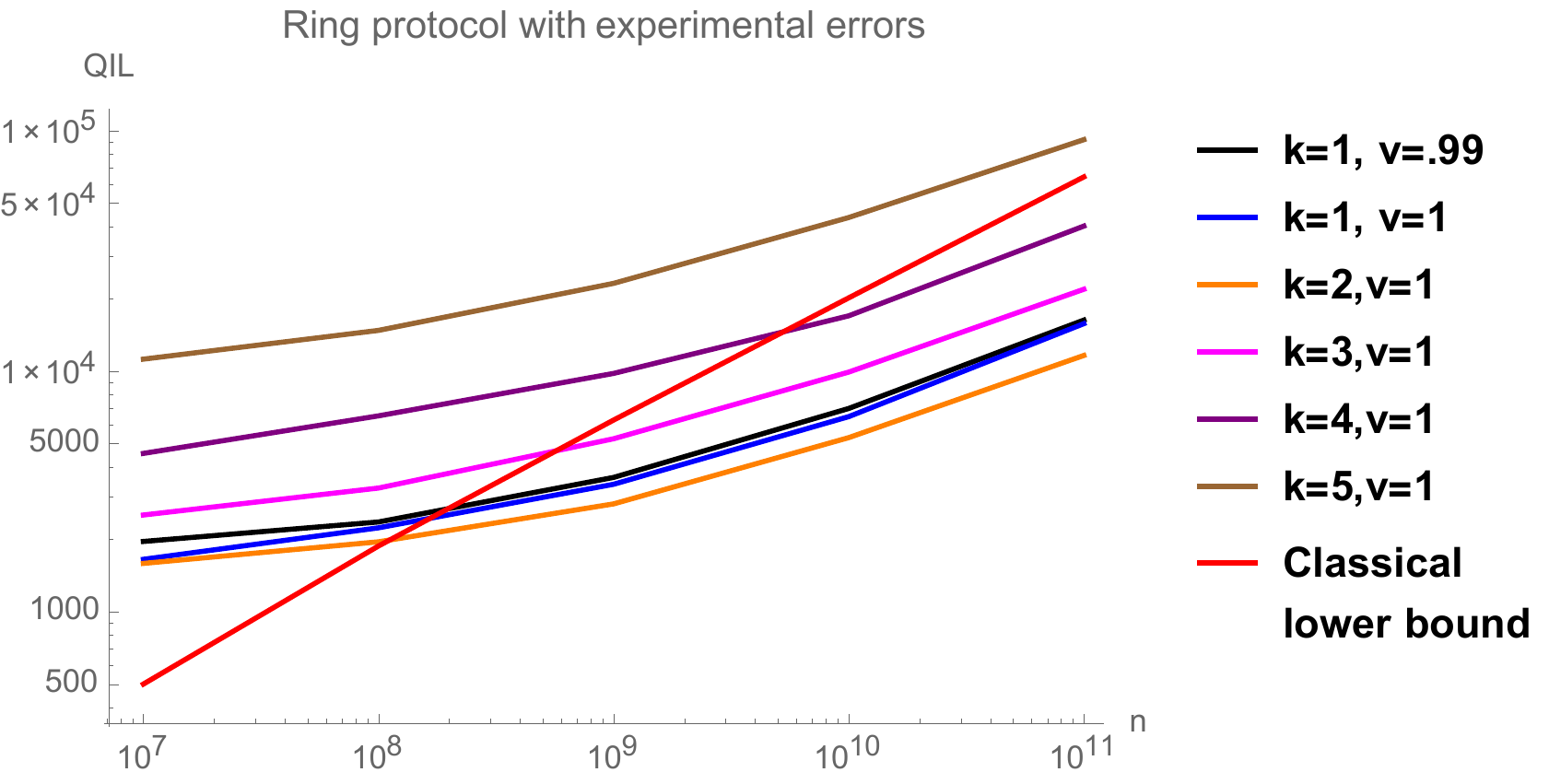}
\caption{Quantum information leakage (QIL), measured in bits, as a function of the input size $n$ for our ring protocols to attain error probability $\epsilon=0.01$ under transmittivity $\eta=0.3$ and dark count probability $p_{\text{dark}}=7.3\times 10^{-11}$. Existing optical protocol with interferometric visibility $99\%$ included for reference. \label{QF_exp}}
\end{figure}

In Figure~\ref{QF_exp} we plot the information leakage of the ring protocols under the bound of Section~\ref{impoptinf} for realistic experimental imperfections, compared to the classical information leakage lower bound of \cite{1607.07516}. For given values of $m$, $k$, $\delta$, and $p_{\text{dark}}$ we have chosen the signal amplitude $\beta_k$ to attain the desired error probability $\epsilon$ under the approximation \eqref{pdk}. As before, we have optimized over $\delta$ and assumed the code $\tE$ saturates the Gilbert-Varshamov bound \eqref{bingv}. This plot uses worst case error probability $\epsilon=0.01$, transmittivity $\eta=0.3$, and dark count probability per signal $p_{\text{dark}}=7.3\times 10^{-11}$. We include a plot of of the existing optical protocol with interferometric visibility $99\%$ for reference. We observe the same hierarchy as the ideal setting, but as dark counts have less effect for protocols sending fewer signals, the $k=2$ (two-bit) protocol now has visibly lower information leakage than the $k=1$ protocol. We have also observed the same hierarchy for $p_{\text{dark}}$ in the range $[0,10^{-9}]$ and $\epsilon$ in the range $[10^{-5},10^{-2}]$. We speculate that the same behaviour holds for the lattice protocol under experimental imperfections.

%%%%%%%%%%%%%%%%%%%%%%%%%%%%%%%%%%%%%%%%%%%%%%%%%%%%%
\section{Information Leakage}\label{infleaksection}
%%%%%%%%%%%%%%%%%%%%%%%%%%%%%%%%%%%%%%%%%%%%%%%%%%%%%

In this section we bound the information leakage of the protocols we have considered. The quantum information leakage of a simultaneous message passing model protocol $\Pi$ in which, conditioned on input $x$, Alice (Bob) sends state $\sigma_x$ to the referee, is given by \cite{5231456}
\begin{align}\label{infleak1}
\text{QIL}(\Pi)=\max_{P\in \Pr(X \times Y)} \text{I}(XY:AB)_{\rho},
\end{align}
where
\begin{align*}
\rho=\sum_{x, y \in \{0,1\}^n} P(x,y) \ketbra{x y}{x y}^{XY}\otimes \sigma_x^A\otimes \sigma_y^B.
\end{align*}
For every protocol we consider, $\sigma_x=\ketbra{\psi_x}{\psi_x}$ is pure, which implies $\text{H}(AB|XY)=0$, so Eq. \eqref{infleak1} reduces to 
\begin{align}\label{infleak}
\text{QIL}(\Pi)=\max_{P\in \Pr(X \times Y)} \tH(AB)_\rho.
\end{align}

To bound the information leakage, we will find an orthonormal basis for the Hilbert space used in each protocol such that the expectation values of $\ketbra{\psi_x}{\psi_x}$ w.r.t. this basis form a fixed probability vector $\Lambda$ for all $x \in \{0,1\}^n$. The Schur-Horn theorem \cite{schur1923uber,10.2307/2372705} then implies $\rho_{AB}\succ \Lambda^{\otimes 2}$, so $\text{H}(AB)\leq \text{H}(\Lambda^{\otimes 2})$ (see, e.g. \cite{nielsen}), which implies
\begin{align}\label{QILbound}
\text{QIL}(\Pi)\leq 2\text{H}(\Lambda)
\end{align}
by the additivity of entropy under tensor product. While not all simultaneous message passing model protocols will have such a basis, we do find such a basis for both the interpolation and optical ring protocol families, and use this technique to bound their information leakage as $\mathcal{O}(\log n)$. In Appendix~\ref{optinf} we bound the information leakage of the optical lattice protocol family as $\mathcal{O}(\log n)$ (for which we were unable to find such a basis). Our Appendix~\ref{optinf} bound also holds for the optical ring protocols, but we have found numerically that our bound of this section is $15-40\%$ lower in the considered parameter regime. This follows intuitively from the fact that the bound in this section takes into account the particular form of the states used, whereas the Appendix~\ref{optinf} bound simply projects onto a neighbourhood of the total mean photon number.

\subsection{Information leakage for family of interpolation protocols}\label{intinf}
%%%%%%%%%%%%%%%%%%%%%%%%%%%%%%%%%%%%%%%%%%%%%%%%%%

Here we bound the information leakage for the family of interpolation protocols under the choice of qubit overlap ${\bkk{q_0^{(k)}}{q_1^{(k)}}}=1-k/m$ as $\mathcal{O}(\log n)$:

\begin{proposition}\label{intinfprop}
For any fixed error probability $\epsilon>0$, there exists a constant $C\geq 0$ such that for every positive integer $n$ the following holds: for all $k=1,\dots,m$ (where $m$ is the length of the codewords $\tE(x)\in \{0,1\}^m$), the information leakage of the interpolation protocol with block size $k$ and qubit overlap ${\bkk{q_0^{(k)}}{q_1^{(k)}}}=1-k/m$ is no greater than $C \log (n)$.
\end{proposition}

As a corollary, for block size $k$ being given by any function of $n$ such that $k(n) \in [m]$ for each $n$, the family of protocols which uses the interpolation family with block size $k(n)$ for each $n$ has information leakage $\mathcal{O}(\log n)$. For example, $k$ could be held to a fixed constant as in the existing optical protocol, or the ratio $k/m$ could be held fixed as in the original protocol.

\begin{proof}
We prove that the information leakage is upper bounded by $C' r \log n$ for some $C'> 0$. In Appendix~\ref{interperr} we show that for fixed error probability $\epsilon$, the repetition number $r$ is fixed, completing the proof.

Note that the states
%with $\kett{q_a^{(k)}}$ replaced by $\sqrt{1-p_k}\ket{00}+\sqrt{p_k}\ket{a}$ for $p_k=1-\abss{\bkk{q_0^{(k)}}{q_1^{(k)}}}$, $a\in\{0,1\}$, because the overlap structure is preserved. The overlap structure of the resulting set of states is exactly reproduced by the states
\begin{flalign*}
\kett{\phi_x^{(k)}}=&\bigotimes_{j=1}^{\cl{m/k}} \Bigg[\sqrt{1-p_k}\ket{00}\nonumber\\
&\hspace{.3in}+\sqrt{\frac{p_k}{2k}}\sum_{i\in \tI[j,k]} \ket{i}\left(\ket{0}+(-1)^{\tE(x)_i}\ket{1}\right)\Bigg]^{}
\end{flalign*}
for $p_k=1-\abss{\bkk{q_0^{(k)}}{q_1^{(k)}}}$ reproduce the overlap structure of the states $\ispa{x}{}{k}$, so by Property 3 of \cite{doi:10.1142/S0219749904000031} (reviewed in Section~\ref{adaptopt}) they are equal up to a change of basis.
%Thus, by Property 3 of \cite{doi:10.1142/S0219749904000031} and transitivity of the change of basis equivalence relation (see Section~\ref{adaptopt}) the states $\kett{\phi_x^{(k,r)}}$ are equal to the states $\ispa{x}{}{k}^{\otimes r}$ up to a change of basis.
For all $x\in \{0,1\}^n$, the diagonal entries of $\kbihpa{x}{}{k}$ are given by
\begin{align}
\Lambda_{\tI, k}^{\otimes \cl{m/k}}=\left(1-p_k,\frac{p_k}{2k},\dots,\frac{p_k}{2k}\right)^{\otimes \cl{m/k}}.
\end{align}
Thus, by Eq. \eqref{QILbound} the information leakage of the interpolation protocol with block size $k$ under the choice $p_k=k/m$ is upper bounded by
\begin{align}
\mathrm{QI}&\mathrm{L}(\Pi^\tI_k)\leq2\cl{m/k} r \text{H}(\Lambda_{\tI,k})\nonumber\\
&=2\cl{m/k} r \left[(1-k/m)\log\left(\frac{1}{1-k/m}\right)+(k/m)\log \left(2m\right)\right] \nonumber\\
&\leq 2r(2+(1+k/m)\log(2m)).
\end{align}
As $\tE$ is a constant rate code, then $m$ is linear in $n$, so this quantity is upper bounded by $C' r \log n$ for some fixed constant $C' > 0$.
\end{proof}

%%%%%%%%%%%%%%%%%%%%%%%%%%%%%%%%%%%%%%%%%%%%%%%%%%
\subsection{Information leakage for family of optical ring protocols}\label{impoptinf}
%%%%%%%%%%%%%%%%%%%%%%%%%%%%%%%%%%%%%%%%%%%%%%%%%%

Here we bound the information leakage for the family of optical ring protocols described in Section~\ref{optideal}.
%In Appendix~\ref{optinf} we use a method similar to that used in Theorem 1 of \cite{PhysRevA.89.062305} to bound the information leakage of both the ring and lattice protocol families as $\mathcal{O}(\log n)$ for fixed block size $k$. We have found numerically that for the ring protocols the bound of this section is lower than that of Appendix~\ref{optinf} by $15-40\%$.
We use a similar technique as in the information leakage bound for the family of interpolation protocols, and write the states in a basis for which the diagonal entries form a fixed probability vector $\Lambda$.

For block size $k$ and total mean photon number $\mu_k$, each signal consists of one of $2^k$ coherent states equally spaced on a ring with amplitude $\beta_k=\sqrt{\mu_k/(m/k)}$. Formally, each signal is contained in the set
\begin{align}
S_k=\left\{\ket{\omega^j\beta_k}\bra{\omega^j\beta_k}, j=0,\dots, 2^k-1 \right\},
\end{align}
where
\begin{align}
\omega=\exp\left[\frac{2\pi i}{2^k}\right].
\end{align}
%Now we show that the set $S_k$ can be written in a basis such that the diagonal entries are equal for all $j=0,\dots, 2^k-1$.
As $\omega^{jl}=\omega^{j n}$ for all $n \equiv l \Mod 2^k$, then the states in $S_k$ are equal to
\begin{align}\label{phasebasis}
\ket{\omega^j \beta_k}=e^{\frac{-\abs{\beta_k}^2}{2}}\sum_{l=0}^{2^k-1} \omega^{jl} \sum_{h \equiv l \Mod 2^k} \frac{\beta_k^h}{\sqrt{h!}}\ket{h}
\end{align}
for $j=0,\dots,2^k-1$, which can be written in a basis as
\begin{align}
\sum_{l=0}^{2^k-1} \omega^{jl} \lambda_l \ket{l}
\end{align}
for
\begin{align}
\lambda_l=e^{\frac{-\abs{\beta_k}^2}{2}}\sqrt{\sum_{h \equiv l \Mod 2^k} \frac{\abs{\beta_k}^{2h}}{h!}}.
\end{align}
In this basis, the diagonal entries of each state in $S_k$ form the probability vector $\Lambda_{\tO,k}=(\lambda_0^2,\dots, \lambda_{2^k-1}^2)$.

Writing each signal of the states used in the optical ring protocol in this basis, for all $x\in\{0,1\}^n$
%note name states?
the diagonal entries of the transformed states are given by $\Lambda_{\tO,k}^{\otimes m/k}$. By Eq. \eqref{QILbound} and additivity of entropy under tensor product, the information leakage of the optical ring protocol $\Pi_k^\tO$ with block size $k$ is upper bounded by

\begin{align}\label{optbound}
\text{QIL}(\Pi_k^\tO)\leq \frac{2 m}{k} \text{H}(\Lambda_{\tO,k}).
\end{align}
This bound is straightforward to calculate in practice, and was used to produce Figures~\ref{QF_ideal} and~\ref{QF_exp}. We have found numerically that this bound gives an advantage of 15-40\% over the bound of Appendix~\ref{optinf} in our considered parameter regime. This bound can also be used to prove the asymptotic information leakage $\mathcal{O}(\log n)$ for any fixed block size $k$ using standard techniques. 
%for large $n$, the signal amplitude $\beta_k=\mu_k/(m/k)$ is small (we show in Appendix~\ref{optenc} that $\mu_k$ is fixed by the desired error probability $\epsilon$ and is independent of $n$). Thus, in equation \eqref{optbound}, apart from the zero and single photon components of $\lambda_0^2$ and $\lambda_1^2$ respectively, all other terms are $\mathcal{O}((1/n)\log(n))$, and thus do not contribute to the asymptotic information leakage. Using this simplification, straightforward calculation reveals that the information leakage is $\mathcal{O}(\log n)$ for any fixed block-size $k$.

%Using this bound, the information leakage for fixed $k$ can be calculated as $\mathcal{O}(\log n)$ using the fact that for large $n$ the signal amplitude $\beta_k=\mu_k/(m/k)$ is small (we show in Appendix~\ref{optenc} that $\mu_k$ is fixed by the error probability $\epsilon$ and is independent of $n$). Thus, in equation \eqref{optbound} all terms besides the zero and single photon components of $c_0^2$ and $c_1^2$ respectively are $\mathcal{O}((1/n)\log(n))$ and can be ignored in the asymptotic analysis.

%%%%%%%%%%%%%%%%%%%%%%%%%%%%%%%%%%%%%%%%%%%%%%%%%%%%%
\section{Conclusion}
%%%%%%%%%%%%%%%%%%%%%%%%%%%%%%%%%%%%%%%%%%%%%%%%%%%%%
In this work we developed several families of quantum fingerprinting protocols. One family demonstrates a trade-off between the number of signals sent and the dimension of each signal. The other families use coherent state signals arranged in a ring and lattice in phase space, thus catering to optical implementations. We found that one such coherent state protocol reduced the number of signals from the existing coherent state protocol by a factor $1/2$, while also reducing the information leakage. Although the other protocols in the ring and lattice families use even fewer signals, we found convincing numerical evidence that they have higher information leakage under the bounds we have used. It would be interesting to investigate whether some other family of coherent state protocols might further reduce the number of signals while maintaining or reducing the information leakage, and whether our information leakage bounds could be improved.

We also proved that a simple beampslitter measurement similar to that used in \cite{andersson2006experimentally} performs optimal unambiguous state comparison between two coherent states given with equal a priori probabilities. It would be interesting to explore whether a similar measurement could be used to perform optimal unambiguous state comparison of coherent states given with arbitrary a priori probabilities.
%\begin{itemize}
%\item Question: Is there a different family of optical protocols for EQ which reduce number of signals while maintaining the transmitted information?
%\end{itemize}
%%%%%%%%%%%%%%%%%%%%%%%%%%%%%%%%%%%%%%%%%%%%%%%%%%%%%
\section{Acknowledgements}
%%%%%%%%%%%%%%%%%%%%%%%%%%%%%%%%%%%%%%%%%%%%%%%%%%%%%
We thank Konrad Banaszek for his insight on the $q-$ary encoding described in Appendix~\ref{optenc}, which helped us improve the presentation of these results. We thank Dave Touchette for his insight on the information leakage bound of Appendix~\ref{asymptoptical}. We thank Juan Miguel Arrazola, Richard Cleve, David Lovitz, and John Watrous for helpful discussions. This research was supported in part by NSERC, Industry Canada and ARL CDQI program. The Institute for Quantum Computing and the Perimeter Institute are supported in part by the Government of Canada and the Province of Ontario.

%We thank Konrad Banaszek for his insight on the $d-$ary encoding of Section III C, which helped us improve the presentation of these results. BL thanks Richard Cleve, David Lovitz, Dave Touchette, and John Watrous for helpful insight and discussions.
\section{Appendix}
\appendix

%%%%%%%%%%%%%%%%%%%%%%%%%%%%%%%%%%%%%%%%%%%%%%%%%%%%%
\section{Beamsplitter measurement in qubit basis, and optimal unambiguous state comparison with a beamsplitter}\label{beamcomp}
%%%%%%%%%%%%%%%%%%%%%%%%%%%%%%%%%%%%%%%%%%%%%%%%%%%%%

Here we explicitly write any set of two coherent states $\{\ket{\beta_0},\ket{\beta_1}\}$ in a basis as qubits, and construct a measurement in this basis (based on a measurement introduced in \cite{BARNETT2003189}) which reproduces the outcome probabilities of the beamsplitter measurement described in Section~\ref{adaptopt}. Using a result of \cite{BARNETT2003189} we then find that a beamsplitter measurement with single photon threshold detectors placed at both output ports performs optimal unambiguous state comparison on $\{\ket{\beta},\ket{-\beta}\}$ for any complex number $\beta$ when the states are given with equal a priori probabilities.

We begin by writing $\{\ket{\beta_0},\ket{\beta_1}\}$ in a basis as qubits. Let $\beta=\frac{1}{2}(\beta_0-\beta_1)$ and
\begin{align}\label{cohtoqubit}
\ket{q_a}&=e^{\frac{-\abs{\beta}^2}{2}}\left[\sqrt{\cos\text{h}(\abs{\beta}^2)}\ket{0}+(-1)^a \sqrt{\sin\text{h}(\abs{\beta}^2)}\ket{1}\right]\nonumber\\
&=\sqrt{1-p}\ket{0}+(-1)^a \sqrt{p}\ket{1}
\end{align}
for $p=\exp[{-\abs{\beta}^2}]\sinh (\abs{\beta}^2)$. It is straightforward to verify that $\abs{\bk{q_0}{q_1}}=\abs{\bk{\beta_0}{\beta_1}}$, so by Property 3 of \cite{doi:10.1142/S0219749904000031} (reviewed in Section~\ref{adaptopt}), the sets $\{\ket{q_0},\ket{q_1}\}$ and $\{\ket{\beta_0},\ket{\beta_1}\}$ are equal up to a change of basis.
%any two non-orthogonal states can be written in a basis as $\{\ket{q_0},\ket{q_1}\}$ for appropriate choice of $\beta$, which can be written in a basis as $\{\ket{\beta},\ket{-\beta}\}$. Thus, not only can any two coherent states be written in a basis as two qubits, but any two states can be written in a basis as coherent states of equal amplitude and opposite phase.

%We now review a known result in unambiguous state discrimination which we will use in the description of the beamsplitter measurement in the qubit basis: given an unknown state from a set of two states $\{\ket{\psi},\ket{\phi}\}$, there exists a measurement which unambiguously determines the identity of the state with probability $1-\abs{\bk{\psi}{\phi}}$. In fact, when the states are given with equal a priori probabilities this measurement maximizes the overall probability of an unambiguous outcome.

Now we review a measurement (introduced in \cite{BARNETT2003189}) in the qubit basis which reproduces the outcome probabilities of the beamsplitter measurement described in Section~\ref{adaptopt}, i.e. on input $\ket{q_a}\ket{q_b}$ for $a,b\in \{0,1\}$, it outputs ``different'' (previously, ``dark port detection'') with probability $1-\abs{\bk{q_a}{q_b}}$, and fails to output ``different'' with probability $\abs{\bk{q_a}{q_b}}$.

The measurement is a direct sum of an unambiguous state discrimination measurement and a controlled-swap measurement on the first and second terms of the decomposition
\begin{align}\label{qq}
\ket{q_a}\ket{q_b}=&(1-p)\ket{00}+(-1)^{a+b} p \ket{11}\nonumber\\
&+\sqrt{p(1-p)}[(-1)^b \ket{01}+(-1)^a \ket{10}].
\end{align}
Operationally, a non-destructive measurement is first performed with two measurement operators, the first (second) of which projects onto the subspace containing the first (second) term of the decomposition \eqref{qq}. If it projects onto the first term, the remaining state takes one of two forms depending on the equality of $a$ and $b$. In this case, an unambiguous state discrimination (USD) measurement is performed to distinguish between these two states. In particular, the USD measurement is performed which is optimal for the case in which each state is given with equal a priori probabilities \cite{PERES198819}. Of course, this choice is sub-optimal for different a priori probabilities, but we make it because it gives rise to a measurement which reproduces the outcome probabilities of the beamsplitter measurement. We refer to the two unambiguous outcomes of this measurement as ``plus'' and ``minus'' corresponding to the two possible states of the first term of Eq. \eqref{qq}, and the inconclusive outcome as ``?''. If the non-destructive measurement projects onto the second term of Eq. \eqref{qq}, a controlled-swap measurement is performed (which effectively projects onto the symmetric and anti-symmetric subspaces). Outcomes ``minus'' and ``anti-symmetric'' are mapped to a single outcome ``different'' which unambiguously determines $a\neq b$, and occurs with probability $1-\abs{\bk{q_a}{q_b}}$ \cite{BARNETT2003189}, thus reproducing the outcome probabilities of the beamsplitter measurement.

Following \cite{BARNETT2003189}, we map outcomes ``plus'' and ``symmetric'' to a single outcome ``same'' which unambiguously determines $a=b$ with probability $1-\abs{\bk{q_a}{q_{\overline{b}}}}$ for $\overline{b}=1\oplus b$ \cite{BARNETT2003189}. In \cite{BARNETT2003189} it is shown that when the states are given with equal a priori probabilities, this measurement performs optimal unambiguous comparison between the cases $a=b$ and $a\neq b$, i.e. it minimizes the probability of an inconclusive outcome ``?''.

%Now we refine the above measurement (following \cite{BARNETT2003189}) to perform optimal unambiguous state comparison, i.e. to unambiguously discriminate between the cases $a=b$ and $a\neq b$, and minimize the probability of an inconclusive outcome. We map outcomes ``minus'' and ``anti-symmetric'' to a single outcome ``different'', and
%outcomes ``plus'' and ``symmetric'' to a single outcome ``same''. Outcome ``different'' unambiguously determines $a\neq b$, and with probability $1-\abs{\bk{q_a}{q_b}}$. Outcome ``same'' unambiguously determines $a=b$ with probability $1-\abs{\bk{q_a}{q_{\overline{b}}}}$ ($\overline{b}=1\oplus b$) \cite{BARNETT2003189}. In \cite{BARNETT2003189} it is shown that when $\Pr(a,b)=1/4$ for all $a,b \in \{0,1\}$ this measurement performs optimal unambiguous comparison between the cases $a=b$ and $a\neq b$, i.e. it minimizes the probability of an inconclusive outcome ``?''.

On input $\ket{(-1)^a\beta}\ket{(-1)^b \beta}$ to a beamsplitter with single photon threshold detectors placed at both the light and dark ports, the dark port registers a detection with probability $1-\abss{\bkk{(-1)^a\beta}{(-1)^{b}\beta}}$ and the light port registers a detection with probability $1-\abss{\bkk{(-1)^a\beta}{(-1)^{\overline{b}}\beta}}$, which are identical to the outcome probabilities of the above optimal unambiguous state comparison measurement in the qubit basis. Thus, this beamsplitter measurement performs optimal unambiguous state comparison on $\{\ket{\beta},\ket{-\beta}\}$ when the states are given with equal a priori probabilities.

\section{Error analysis}
%%%%%%%%%%%%%%%%%%%%%%%%%%%%%%%%%%%%%%%%%%%%%%%%%%%%%

In Appendix~\ref{interperr} we calculate the worst case error probabilities for the family of interpolation protocols, which we use in Section~\ref{intinf} to show that they have information leakage $\mathcal{O}(\log n)$.

In Appendix~\ref{optenc} we calculate the worst case error probabilities for the family of optical ring protocols described in Section~\ref{opt_eq_families}, which we use in Section~\ref{impoptinf} and Appendix~\ref{optinf} to show that they have information leakage $\mathcal{O}(\log n)$. We also prove optimality of these protocols over several similar protocols which use coherent state signals arranged in a ring in phase space, and state without proof analogous results for optical lattice protocols.

In Appendix~\ref{opmeas} we determine the optimal one-sided error measurement for any simultaneous message passing model equality protocol, and show that the worst case error probability is lower bounded by the square of the error probability of the beamsplitter measurement. We use these results in Section~\ref{optical} to show numerically that larger block sizes $k$ do not outperform the two-bit protocol, even under the optimal measurement.

%In all protocols that we consider, to make different states sufficiently distinguishable to Ray, inputs $x, y \in \{0,1\}^n$ are mapped to codewords $\tE(x),\tE(y)\in \{0,1\}^m$ with minimum distance $\delta$, which are subsequently encoded into states whose overlap is a decreasing function of the distance between codewords. We also briefly consider $d-$ary codewords in Appendix~\ref{optenc}.
For most protocols that we consider, each signal encodes several bits of $\tE(x)\in \{0,1\}^m$. For different inputs $x\neq y \in\{0,1\}^n$, the error probability depends on the distribution of the bit differences between the codewords across the signals. For a given code, the worst case distribution over the particular $2^n$ codewords could be difficult to calculate. Instead, for most arguments we make the following simplifying assumption.
\begin{remark}\label{worstcase}
In calculating the worst case error probability of a protocol, we assume that the code is uncharacterized apart from its minimum distance, and take the worst case over all strings $\tE_A \neq \tE_B$ in the output space of the code which differ by at least the minimum distance.
\end{remark}

%%%%%%%%%%%%%%%%%%%%%%%%%%%%%%%%%%%%%%%%%%%%%%%%%%%%%
\subsection{Error analysis for family of interpolation protocols}\label{interperr}
%%%%%%%%%%%%%%%%%%%%%%%%%%%%%%%%%%%%%%%%%%%%%%%%%%%%%

Here we calculate the worst case error probability of the interpolation protocol with block size $k$ for any choice of positive qubit overlap ${\bkk{q_0^{(k)}}{q_1^{(k)}}}\geq 0$. Similar results hold for arbitrary overlap, but we make this assumption because it simplifies the algebra. We then show that under the choice ${\bkk{q_0^{(k)}}{q_1^{(k)}}}=1-k/m$ the error probability is upper bounded by $2^{-\delta r}$, for repetition number $r$ denoting the number of identical copies of $\ispa{x}{}{k}$ sent.
%note repetition number r defined?
This implies that for a given desired error probability $\epsilon$, the repetition number can be fixed independent of $n$. We use this fact in the proof of Proposition~\ref{intinfprop} that the information leakage of the interpolation is $\mathcal{O}(\log n)$.

As mentioned in Remark~\ref{worstcase}, the worst case is taken over all strings $\tE_A \neq \tE_B \in \{0,1\}^m$ which differ by minimum distance $m \delta$ bits. To determine the worst case over this set, we first calculate the probability that the referee's measurement on the $j$-th pair of signals of $\ispa{xy}{}{k}$ returns ``dark port detection'' or ``anti-symmetric'' under the assumption that $\tE_{A,i}\neq \tE_{B,i}$ for $d$ indices $i \in \tI[j,k]$. Recall that the qubit-basis beamsplitter measurement is performed on the first term of the decomposition \eqref{instatesdecomp}, and the controlled-swap measurement is performed on the second term.

Recall that on input $\kett{q^{(k)}_{0}}\kett{q^{(k)}_{1}}$, the qubit-basis beamsplitter measurement outputs ``dark port detection'' with probability $1-{\bkk{q_0^{(k)}}{q_1^{(k)}}}$ (see Eq. \eqref{beamerr} and Appendix~\ref{beamcomp}). Thus, when $d$ bits differ the qubit-basis beamsplitter measurement on the first term of the decomposition \eqref{instatesdecomp} outputs ``dark port detection'' (``D'') with probability
\begin{align}\label{minusprob}
{\Pr}_k^\text{I}(\text{``D''}| d \text{ bits differ})=\frac{d}{k^2}\left(1-{\bkk{q_0^{(k)}}{q_1^{(k)}}}\right).
\end{align}
Recall that on input $\ket{\psi}$, the controlled-swap outputs ``anti-symmetric'' with probability $\Vert \frac{1}{2} (\mathds{1}-\mathrm{W}) \ket{\psi}\Vert^2$, where $\mathrm{W}$ is the operator which swaps Alice and Bob's systems. It follows that the controlled-swap on the second term of the decomposition \eqref{instatesdecomp} outputs ``anti-symmetric'' with probability
\begin{flalign*}\label{asprob}
{\Pr}_k^\text{I}(&\text{``A-S''}|d)=\frac{1}{4k^2}\sum_{\substack{i,l \in \tI[j,k] \\ i\neq l}} \nsq{\kett{q^{(k)}_{\tE(x)_i}}\kett{q^{(k)}_{\tE(y)_l}}-\kett{q^{(k)}_{\tE(y)_i}}\kett{q^{(k)}_{\tE(x)_l}}}\nonumber\\
&=\frac{1}{2k^2}\Bigg[k(k-1)-\sum_{\substack{i,l \in \tI[j,k] \\ i\neq l}} \bkk{q^{(k)}_{\tE(x)_i}}{q^{(k)}_{\tE(y)_i}}\bkk{q^{(k)}_{\tE(y)_l}}{q^{(k)}_{\tE(x)_l}}\Bigg]\\
&=\frac{1}{2k^2}\Bigg[k(k-1)-(k-d)(k-d-1)\nonumber\\
&\hspace{.6in}-2d(k-d)\bkk{q_0^{(k)}}{q_1^{(k)}}-d(d-1){\bkk{q_0^{(k)}}{q_1^{(k)}}}^2\Bigg]\\
\end{flalign*}
The first and second equalities are straightforward. The third follows from the fact that $(k-d)(k-d-1)$ terms in the sum are equal to one, $2d(k-d)$ terms have one bit different, and $d(d-1)$ terms have two bits different. Let ``no detection'' (``ND'') denote the event that neither ``dark port detection'' nor ``anti-symmetric'' occur for a given signal. After simplification, the probability ${\Pr}_k^\text{I}(\text{``ND''}|d)$ of outcome ``no detection'' when the two $k$-bit blocks differ by $d$ bits is given by
\begin{align}
{\Pr}_k^\text{I}(\text{``ND''}|d)&=1-{\Pr}_k^\text{I}(\text{``D''}| d)-{\Pr}_k^\text{I}(\text{``A-S''}|d)\nonumber\\
&=1-\frac{d}{k}p_k\left(1-\frac{(d-1)p_k}{2k}\right),
\end{align}
where $p_k=1-\bkk{q_0^{(k)}}{q_1^{(k)}}$. For a given distribution of bit differences $d_1,\dots,d_{m/k}$ beetween two codewords among the $m/k$ blocks, the total error probability for repetition number $r$ is given by
\begin{align}\label{intermedierr}
{\Pr}_k^\text{I}(\text{Err}|d_1,\dots,d_{m/k})=\prod_{i=1}^{m/k} {\Pr}_k^\tI(\text{``ND''} | d_i)^r,
\end{align}
which reproduces Eq. \eqref{orpsame} for $p_k=1$ and $k=m$ in the worst case $d=m \delta$, as expected.

Now we prove that the worst case error probability for each block size $k$ occurs when all bit differences between the codewords are consolidated into the fewest number of signals. Indeed, by straightforward calculation it can be shown that for all $0\leq p_k \leq1$, for every pair of integers $1\leq d, c \leq k-1$ such that $d \geq c$,
\begin{align*}
{\Pr}_k^\text{I}(\text{``ND''}|d){\Pr}_k^\text{I}(\text{``ND''}|c)\leq {\Pr}_k^\text{I}(\text{``ND''}|d+1){\Pr}_k^\text{I}(\text{``ND''}|c-1),
\end{align*}
which proves the claim. Thus, the worst case error probability is given by
\begin{align}\label{err}
&{\Pr}^\tI_k(\text{Err})=\nonumber\\
&\left[1-p_k\left(1-\frac{(k-1)p_k}{2k}\right)\right]^{\fl{\frac{m \delta}{k}}r}\left[1-p_k\frac{t}{k}\left(1-\frac{(t-1)p_k}{2k}\right)\right]^r
\end{align}
for $t$ the remainder given by $\frac{m \delta}{k}=\fl{\frac{m \delta}{k}}+\frac{t}{k}$.

In what follows, we show that under the choice $p_k=k/m$ the worst case error proability is upper bounded by a constant to the power $r$. This implies that any fixed error $\epsilon$ can be attained with fixed repetition number. We use this fact in Proposition~\ref{intinfprop} to show that the information leakage of the interpolation under this choice is $\mathcal{O}(\log n)$.

Under the choice $p_k=k/m$, the worst case error probability is upper bounded by $2^{-\delta r}$:
\begin{align}\label{iner}
&{\Pr}^\tI_k(\text{Err}|p_k=k/{m})=\nonumber\\
&\left[1-\frac{k}{m}\left(1-\frac{(k-1)}{2m}\right)\right]^{\fl{\frac{m \delta}{k}}r}\left[1-\frac{t}{m}\left(1-\frac{(t-1)}{2m}\right)\right]^r\nonumber\\
&\leq \left[1-\frac{k}{m}\left(1-\frac{k}{2m}\right)\right]^{\fl{\frac{m \delta}{k}}r}\left[1-\frac{t}{m}\left(1-\frac{t}{2m}\right)\right]^r\nonumber\\
&\leq \left[1-\frac{k}{m}\left(1-\frac{k}{2m}\right)\right]^{\frac{m \delta }{k}r}\nonumber\\
&\leq 2^{-\delta r}
\end{align}
%\begin{align}\label{iner}
%\left[\frac{1}{2}\left(1+(1-p_k)^2-\frac{p_k^2}{k}\right)\right]^\fl{\frac{m \delta}{k}}\left[1-\frac{rp_k}{k}\left(1-\frac{(r-1)p_k}{2k}\right)\right],
%\end{align}
for all $1 \leq k \leq m$. The equality follows from simplification of Eq. \eqref{err}. The first inequality is straightforward, and the second and third both follow from the fact that the function $[1-x(1-\frac{x}{2})]^{1/x}$
%\begin{align}
%\left[1-x\left(1-\frac{x}{2}\right)\right]^{1/x}
%\end{align}
is strictly increasing with $x$ for all $0<x< 1$.

%it can be shown that for any fixed $0<\delta<1/2$ the quantity \eqref{iner} is less than a fixed constant that is less than 1 and independent of both $k$ and $m$.

%%%%%%%%%%%%%%%%%%%%%%%%%%%%%%%%%%%%%%%%%%%%%%%%%%%%%%%
\subsection{Error analysis for family of optical protocols}\label{optenc}
%%%%%%%%%%%%%%%%%%%%%%%%%%%%%%%%%%%%%%%%%%%%%%%%%%%%%%%
Here we derive the worst case error probability of the optical ring protocol described in Section~\ref{opt_eq_families} with block size $k$, which we use in Section~\ref{impoptinf} and Appendix~\ref{optinf} to bound the information leakage as $\mathcal{O}(\log n)$. We also prove that for all $k=1,\dots,4$ the ring Gray code is an optimal encoding of $k$-bit blocks of binary codewords into coherent states arranged in a ring in phase space. We then show that the ring Gray code outperforms an alternative which uses $q-$ary codewords. We state without proof analogous results for the optical lattice protocols.
\begin{lemma}\label{ring_worst_case}
For any positive integer $k$, the ideal ring protocol described in Section~\ref{optideal} with block size $k$, using the Gray code, states of total mean photon number $\mu_k$, and the beamsplitter measurement described in Section~\ref{adaptopt}, satisfies the following.

The worst case error probability ${\Pr}_k^\tO(\mathrm{Err})$ is upper bounded by
\begin{align}\label{ring_worst_case_err}
\exp\bigg[&-\mu_k \bigg[1-(1-(k \delta- \fl{k \delta})) \cos\left(\frac{2\pi \fl{k \delta}}{2^k}\right)\nonumber\\
&\hspace{.3in}-(k \delta-\fl{k \delta})\cos\left(\frac{2\pi (\fl{k \delta}+1)}{2^k}\right)\bigg]\bigg].
\end{align}
Furthermore, if the worst case error probability is taken over all codewords $\tE_A\neq \tE_B \in \{0,1\}^m$ which differ by at least the minimum distance $m \delta$ bits (as described in Remark~\ref{worstcase}), this bound is attained with equality for all $0\leq \delta \leq 3/k$.
%Furthermore, \eqref{ring_worst_case_err} is the worst case error probability of any encoding of $k$-bit blocks into equally spaced nodes on a ring for which nearest neighbour nodes differ in one bit.
\end{lemma}

As a corollary, since every binary code with more than two codewords has minimum distance $\delta\leq 1/2$, then for all $k=1,\dots, 6$, the worst case error probability ${\Pr}_k^\tO(\mathrm{Err})$ is given by Eq. \eqref{ring_worst_case_err} with equality in the ideal setting.

\begin{proof}
The error probability of the beamsplitter measurement is given by the probability that ``no dark port detection'' (``ND'') occurs for every pair of signals. The worst case error probability depends on the worst case distribution of the bit differences between the codewords across the signals. To upper bound the worst case error probability, we first bound the probability ${\Pr}_k^\tO(\text{``ND''} | d)$ of ``ND'' when a pair of $k$-bit blocks differ by $d$ bits. As a property of the Gray code, nearest-neighbour coherent state signals in phase space correspond to blocks which differ by one bit. Thus, for any pair of blocks which differ by $d$ bits, the corresponding pair of coherent state signals must be spaced at least $d$ steps apart on the ring.
%note mention d<2^{k-1} step?
It follows that ${\Pr}_k^\tO(\text{``ND''} | d)$ is upper-bounded by the probability of ``ND'' when the pair of coherent state signals are spaced $d$ steps apart on the ring, which is given by
% straightforward to show using equation~\eqref{beamerr} that the probability of ``ND'' is given by the righthand side of the expression
%Recall equation \eqref{beamerr}, which gives the probability of ``ND'' given two coherent state inputs.
\begin{align}\label{p_NC_d}
{\Pr}_k^\tO(\text{``ND''} | d)\leq \exp\left[{-\abs{\beta_k}^2 \left(1-\cos\left(\frac{2\pi d}{2^k}\right)\right)}\right],
\end{align}
which follows from straightforward calculation using equation~\ref{beamerr} (recall $\beta_k=\sqrt{\mu_k/(m/k)}$ is the signal amplitude). For $d=1,2,3$ under the Gray code, Eq. \eqref{p_NC_d} is satisfied with equality (see Figure~\ref{psk_figure} or \cite{gray1953pulse}).

Now we upper bound the worst case error probability using the bound \eqref{p_NC_d} for each signal. For a given distribution of bit differences $d_1, \dots, d_{m/k}$ among the blocks, the error probability is given by
\begin{align}\label{opterrbound}
{\Pr}_k^\tO(\text{Err}|d_1,&\dots,d_{m/k})=\prod_{i=1}^{m/k} {\Pr}_k^\tO(\text{``ND''} | d_i)\nonumber\\
&\leq \exp\left[{-\mu_k + \abs{\beta_k}^2 \sum_{i=1}^{m/k} \cos\left(\frac{2\pi d_i}{2^k}\right)}\right].
\end{align}
Note that
\begin{flalign*}
&\cos\left(\frac{2\pi d}{2^k}\right)+\cos\left(\frac{2\pi c}{2^k}\right)\nonumber\\ 
&\hspace{.1in} \leq \cos\left(\frac{2\pi}{2^k}\fl{\frac{d+c}{2}}\right)+\cos\left(\frac{2\pi}{2^k} \cl{\frac{d+c}{2}}\right)
%&\geq d\left(1-\cos\left(\frac{2\pi }{2^k}\right)\right)
\end{flalign*}
for all $1\leq d, c \leq k$. This can be proven by direct calculation for $k=3$, and for $k\geq 4$ it follows from the fact that the function $\cos(2\pi \nu)+\cos(2\pi(a-\nu))$ is strictly decreasing with $\nu$ whenever $0\leq \nu < a-\nu < \frac{1}{4}$, along with the fact that $\frac{k}{2^k}\leq \frac{1}{4}$.

It follows that the righthand side of Eq. \eqref{opterrbound} is maximized for strings $\tE_A\neq \tE_B \in \{0,1\}^m$ such that the bit differences $d_1, \dots, d_{m/k}$ are evenly distributed among the $m/k$ blocks and differ by minimum distance $m \delta$ bits. In this case, $\frac{m}{k}(k\delta-\fl{k \delta})$ pairs of signals differ by $\fl{k \delta}+1$ bits, and $\frac{m}{k}(1-(k\delta-\fl{k \delta}))$ pairs differ by $\fl{k \delta}$ bits, giving rise to the bound \eqref{ring_worst_case_err}. Below, we refer to this case of inputs as \textit{adjacent}.

If $\delta \leq 3/k$, then $m \delta \leq 3 m/k$, so when the $m \delta$ bit differences are evenly distributed among the $m/k$ blocks, $d_i \leq 3$ for all $i=1,\dots, m/k$. As mentioned above, under the Gray code such inputs satisfy Eq. \eqref{p_NC_d} with equality for every signal, and thus also satisfy Eq. \eqref{ring_worst_case_err} with equality.
\end{proof}

In the experimental setting, it can be shown using similar techniques to those used in the proof of  Lemma~\ref{ring_worst_case_err} that adjacent inputs give rise to the lowest expected number of outcomes ``dark port detection'' in our considered parameter regime. The approximation \ref{pdk} to the behaviour of the random variable $D_{D,k}$ follows using standard techniques.

Now we prove statements of optimality for the Gray code in the optical ring protocols. Analogous results also hold for the lattice protocols.
\begin{proposition}
In all protocols considered below, assume the referee uses the beamsplitter measurement described in Section~\ref{adaptopt}, and that the worst case error probability is taken over all codewords $\tE_A\neq \tE_B \in \{0,1\}^m$ which differ by at least the minimum distance $m \delta$ bits, as described in Remark~\ref{worstcase}.

For any positive integer $k$, the ideal ring protocol described in Section~\ref{optideal} with block size $k$, using states of total mean photon number $\mu_k$, satisfies the following: For all $0\leq \delta \leq 2/k$, the Gray code minimizes the worst case error probability over all encodings of $k$-bit blocks of binary codewords into a ring of equally spaced coherent state signals in phase space.
\end{proposition}
As a corollary, since every binary code with more than two codewords has minimum distance $\delta\leq 1/2$, the Gray code is optimal for all $k=1,\dots, 4$.
\begin{proof}
We prove optimality of any encoding in which nearest neighbour coherent state signals differ by one bit (of which the Gray code is an example). First we show that for any such encoding, the worst case error probability is given by Eq. \eqref{ring_worst_case_err} with equality. For any such encoding, any pair of coherent state signals which are second-nearest neighbours correspond to a pair of blocks which differ by two bits, and thus satisfy Eq. \eqref{p_NC_d} with equality. 
% there exist $k-$bit blocks of bits corresponding to nodes which satisfy \eqref{p_NC_d} with equality for $d=1,2$.
By the same arguments used to prove Lemma~\ref{ring_worst_case}, it follows that the worst case error probability for all $\delta \leq 2/k$ under any such encoding is given by Eq. \eqref{ring_worst_case_err} with equality.

Now we show that any encoding for which there exist nearest-neighbour coherent state signals which differ by $d'\geq 2$ bits has worst case error probability greater than Eq. \eqref{ring_worst_case_err}. The case assumptions $\delta \leq 2/k$ and $d'\geq 2$ imply $m \delta \leq d' m/k$. Thus, there exist codewords which differ in $m \delta$ bits and for which the bit differences $d_1,\dots,d_{m/k}$ are distributed among the $m/k$ blocks such that $d_i=d'$ for $m \delta/d'$ indices $i$, and all other bit differences are zero. Furthermore, the codewords can be chosen so that for every index $i$ satisfying $d_i=d'$, the corresponding pair of coherent state signals are nearest neighbours in phase space. Such codewords clearly give rise to error probability greater than Eq. \eqref{ring_worst_case_err}. Here we have assumed $d'$ divides $m \delta$, but similar techniques can be used to prove the same statement in the case when $d'$ does not divide $m \delta$.
\end{proof}

We have shown that under certain conditions, the Gray code is an optimal encoding of binary codewords into a ring. We now consider another family of optical ring protocols which map $q$-ary codewords into $q$ equally-spaced nodes on a ring. Under the assumption that all codes saturate the Gilbert -Varshamov bound, we show that the Gray coding of binary codewords outperforms this family for all $q$ powers of two. An analogous result holds for the lattice protocols.

\begin{proposition}
In all protocols considered below, assume all codes saturate the Gilbert-Varshamov bound. Furthermore, assume that the referee uses the beamsplitter measurement described in Section~\ref{adaptopt}, and that the worst case error probability is taken over all codewords $\tE_A \neq \tE_B$ in the output space of the code which differ by at least the minimum distance of the code, as described in Remark~\ref{worstcase}.

Let $q$ be any power of two. For all $\epsilon>0$, for any $q$-ary ring protocol described above which attains error probability $\epsilon$ with total mean photon number $\mu_q$ and $m_q$ signals, the ring protocol described in Section~\ref{opt_eq_families} with block size $k=\log q$ and Gray coding can attain the same error probability $\epsilon$ with the same total mean photon number $\mu_q$ and fewer than $m_q$ signals.
\end{proposition}

As a corollary, because the ring protocol with block size $k=\log q$ uses the same total mean photon number and fewer signals than the $q$-ary ring protocol, then it has lower information leakage under the bound derived in Appendix~\ref{optinf}. It can also be shown that this statement holds under the stronger bound derived in Section~\ref{impoptinf} when $n \gg \mu_q$ using standard approximation techniques.

\begin{proof}
Let $m_q$ denote the length of the $q-$ary code (i.e. the number of ``qits'' in the code), and let $\delta_q m_q$ denote its minimum distance (i.e. the minimum number of differing qits between codewords). It is easy to see that the worst case error probability occurs when the codewords $\tE_A\neq \tE_B\in [q]^{m_q}$ differ by $\delta_q m_q$ qits, and every pair of differing qits correspond to nearest-neighbour coherent state signals; and is given by
\begin{align}\label{errdary}
{\Pr}_q(\mathrm{Err})=\exp\left[-\mu_q \delta_q \left(1-\cos\left(\frac{2\pi}{q}\right)\right)\right].
\end{align}
As the $q-$ary code saturates the ($q-$ary) Gilbert-Varshamov bound, the quantity $\delta_q$ satisfies
\begin{align}\label{dary_rate}
\frac{n}{m_q} = \log q-\delta_q \log(q-1)-\text{h}(\delta_q)
\end{align}
and $0\leq \delta_q<1-1/q$ \cite{van2013introduction}. Note that Eq. \eqref{dary_rate} is the ratio of $n$ to the number of signals used.

Now consider the ring protocol with block size $k=\log q$ using the Gray code, the same total mean photon number, the beamsplitter measurement, and minimum distance 
\begin{align}\label{eq}
\delta=\frac{\delta_q}{\log q}.
\end{align}
Note that
\begin{align}\label{ineq}
\delta < \frac{1-\frac{1}{q}}{\log q}< \frac{2}{\log q}.
\end{align}
The first inequality follows from $\delta_q<1-1/q$ and the second is straightforward. By Lemma~\ref{ring_worst_case}, the inequality \eqref{ineq} implies that the worst case error probability is given by Eq. \eqref{errdary} with equality. As the code saturates the Gilbert-Varshamov bound, the ratio of $n$ to the number signals is given by
\begin{align}\label{gray_rate}
\frac{n}{m/\log q}=(1-\text{h}(\delta_q/\log q)) \log q.
\end{align}
For all $0\leq \delta_q <1-1/q$, we now show that the righthand side of Eq. \eqref{dary_rate} is no greater than the righthand side of Eq. \eqref{gray_rate}, which implies that this protocol sends fewer signals than the $q$-ary ring protocol, completing the proof.

After substituting $\log q=k$ and $\delta_q=k \delta$, the desired inequality becomes
\begin{align}\label{ineqsimp}
\frac{\text{h}(\delta)}{\delta}-\frac{\text{h}(k \delta)}{k \delta}\leq \log(2^k-1)
\end{align}
for all $0\leq \delta \leq (1-2^{-k})/k$. Using standard calculus techniques, it can be shown that the lefthand side of Eq. \eqref{ineqsimp} is a strictly increasing function of $\delta$. Thus, the inequality need only be shown for $\delta=(1-2^{-k})/k$, which is proven using standard techniques.
\end{proof}

%In the experimental setting, loss will not change this behaviour. Furthermore, with fewer signals sent, the effects of dark counts and limited visibility on the error probability decrease, so these imperfections will also not change this behaviour.

%%%%%%%%%%%%%%%%%%%%%%%%%%%%%%%%%%%%%%%%%%%%%%%%%%%%%%%%
\subsection{Optimal measurement}\label{opmeas}
%%%%%%%%%%%%%%%%%%%%%%%%%%%%%%%%%%%%%%%%%%%%%%%%%%%%%%%%
Here we derive the optimal one-sided error measurement for any simultaneous message passing model equality protocol. We then show that the error probability of the optimal measurement is lower bounded by the square of the error probability of the beamsplitter measurement. We use these results in Section~\ref{optical} to show numerically that the optical protocols with block size $k>2$ do not outperform the two-bit protocol, even under the optimal measurement.

%Consider the fo general setting which contains all SMP model equality protocols as special cases.
%In the equality setting, conditioned on inputs $x, y \in \{0,1\}^n$, Alice and Bob send states $\ket{\psi_x}_A$ and $\ket{\phi_{y}}_B$ to Ray (for all protocols that we consider, $\ket{\psi_x}=\ket{\phi_x}$). Ray's optimal measurement minimizes the worst case probability to ouput Equal over all inputs $x \neq y$ under the constraint that for all inputs $x = y$ it never outputs NotEqual.
Consider a general setting in which the referee receives a state $\sigma^{AB}_{z}$, where $z$ is contained in one of two sets EQ or NEQ. The referee then wishes to determine whether $z$ is contained EQ or NEQ under the constraint that if $z \in \text{EQ}$ they never err. It is straightforward to show that the measurement  $\{\Pi_{\text{EQ}}, \mathds{1}-\Pi_{\text{EQ}}\}$ (the projection onto the space spanned by the image of the states $\sigma^{AB}_{z\in \text{EQ}}$ and its orthogonal complement) minimizes the worst case error probability of this task.
%
%
%Consider a general setting in which the referee receives a state $\rho^{AB}_{z}$ for $z \in Z$ chosen according to some probability distribution $P\in \Pr(Z)$. For some partition $Z=\text{EQ}\sqcup \text{NEQ}$, the referee wishes to determine whether $z$ is contained EQ or NEQ under the constraint that if $z \in \text{EQ}$ they never err. Assuming $p(z)>0$ for all $z \in \text{EQ}$, it is straightforward to show that the measurement  $\{\Pi_{\text{EQ}}, \mathds{1}-\Pi_{\text{EQ}}\}$ (the projection onto the space spanned by the image of the states $\rho^{AB}_{z\in \text{EQ}}$ and its orthogonal complement) minimizes the worst case error probability of this task.

\begin{proposition}
In the setting described above, for
\begin{align}\label{zeqneq}
%Z&=\{(x,y) : x,y\in\{0,1\}^n\}\nonumber\\
\mathrm{EQ}&=\{(x,x): x \in\{0,1\}^n\}\nonumber\\
\mathrm{NEQ}&=\{(x,y): x\neq y\in\{0,1\}^n\},
\end{align}
if
\begin{align}
\sigma^{AB}_{(x,y)}=\ketbra{\psi_{x}}{\psi_{x}}\otimes \ketbra{\psi_{y}}{\psi_{y}}
\end{align}
for all $x,y \in \{0,1\}^n$, then the error probability on input $x \neq y \in \{0,1\}^n$ is lower bounded by $\abs{\bk{\psi_x}{\psi_y}}^2$.
\end{proposition}
As a corollary, since the error probability of the beamsplitter measurement in the optical protocols is given by $\abss{\bkk{\psi_x}{\psi_y}}$ on coherent state inputs (see Eq. \eqref{beamerr} and the subsequent discussion), then the error probability of the optimal measurement is lower bounded by the square of the error probability of the beamsplitter measurement.
\begin{proof}
The error probability is lower bounded by
\begin{align*}
\bra{\psi_x \psi_y}\Pi_{\text{EQ}}\ket{\psi_x \psi_y}\geq \frac{2 \abs{\bk{\psi_x}{\psi_y}}^2}{1+ \abs{\bk{\psi_x}{\psi_y}}^2}\geq\abs{\bk{\psi_x}{\psi_y}}^2.
\end{align*}
The second inequality is straightforward and the first is derived by considering only the first two terms of the decomposition 
\begin{align}
\Pi_{\text{EQ}}=\ketbra{\psi_x \psi_x}{\psi_x \psi_x}+\ketbra{\phi}{\phi}+\dots,
\end{align}
where $\ket{\phi}$ is the normalized component of $\ket{\psi_y \psi_y}$ orthogonal to $\ket{\psi_x \psi_x}$.
\end{proof}

%%%%%%%%%%%%%%%%%%%%%%%%%%%%%%%%%%%%%%%%%%%%%%%%%%%%%
\section{Information leakage for families of optical ring and lattice protocols.}\label{optinf}
%%%%%%%%%%%%%%%%%%%%%%%%%%%%%%%%%%%%%%%%%%%%%%%%%%%%%

Here we bound the information leakage of any pure state simultaneous message passing model protocol in which every state $\ket{\psi_{xy}}$ is a tensor product of $m_{k}$ coherent states with total mean photon number lying in a fixed range $[\mumin, \mumax]$ for all $x,y\in\{0,1\}^n$. In Section~\ref{optideal} we use these results to bound the information leakage of the families of optical ring and lattice protocols described as $\mathcal{O}(\log n)$.

%For simplicity of presentation, the index set $\{0,1\}^n$ has been chosen in accordance with the quantum fingerprinting setting, framework, however, the derived bounds hold for any SMP protocol (and corresponding index set) that satisfies the above conditions on the states sent to the referee. In particular, they hold for the Euclidean distance protocol described in Section~\ref{imp_opt}.

In Appendix~\ref{practical} we give a practical bound on the information leakage using a continuity bound on entropy. Due to the dimension dependence of the continuity bound, this bound does not give the desired $\mathcal{O}(\log m_k)$ limiting behaviour, but has the advantage of being straightforward to calculate in practice. In Appendix~\ref{asymptoptical} we bound the asymptotic behaviour as $\mathcal{O}(\log m_k)$.

\subsection{Practical information leakage bound}\label{practical}
Here we use an extension of Theorem 1 of \cite{PhysRevA.89.062305} and a continuity bound on entropy to bound the information leakage of any simultaneous message passing model protocol satisfying the conditions outlined above.

By Eq. \eqref{infleak}, the information leakage is equal to the entropy of $\rho^{AB}$, maximized over prior distributions $P\in \Pr(X\times Y)$. We use the Fannes-Audenaert inequality
\begin{align}
\tH(\rho^{AB}) \leq \tH(\Gamma^{AB}) + \gamma \log \dim (AB)+\text{h}(\gamma),
\end{align}
which bounds $\tH(\rho^{AB})$ in terms of $\tH(\Gamma^{AB})$, $\gamma=\frac{1}{2}\left\Vert \rho^{AB}-\Gamma^{AB} \right\Vert_1$, and $\dim(AB)$ for any state $\Gamma^{AB}$ \cite{Fannes1973,1751-8121-40-28-S18}. We choose $\Gamma^{AB}=\Pi_0 \rho^{AB} \Pi_0/\text{Tr}(\Pi_0 \rho^{AB})$, where $\Pi_0$ projects onto a ``typical subspace'' of $\rho$, given by the set of Fock states of total photon number lying within some radius $\Delta\in \mathds{N}$ of the interval $[\mumin, \mumax]$, as in \cite{PhysRevA.89.062305}. By straightforward extension of Theorem 1 of \cite{PhysRevA.89.062305},

% We then bound $\tH(\sigma)$ using \eqref{pi0dim} and the dimension bound on entropy.

% the states $\kett{\psi_{xy}}$ are $\epsilon'$-close to states of $\mathcal{O}(\log m_k)$ qubits,
%which sta: if every state $\ket{\psi_{xy}}$ is a tensor product of $m_k$ coherent states with total mean photon number lying in a fixed range $[\mumin, \mumax]$, then
%which are obtained by projecting $\kett{\psi_{xy}}$ onto a neighbourhood of the range of total mean photon numbers $[\mumin, \mumax]$. Formally, 
%there exists $\Delta \in \mathds{N}$ such that the projection $\Pi_0$ onto Fock states of total photon number lying within a radius $\Delta$ of the interval $[\mumin, \mumax]$ satisfies
\begin{align}\label{typrob}
\bra{\psi_{xy}}\Pi_0\ket{\psi_{xy}}
&\geq 1-\max\left\{0, e^{-\mumin}\left(\frac{e \mumin}{\mumin-\Delta}\right)^{\mumin-\Delta} \right\}\nonumber\\
&\hspace{.5in}-e^{-\mumax}\left(\frac{e \mumax}{\mumax+\Delta}\right)^{\mumax+\Delta},
%\geq 1-\epsilon'
\end{align}
and
\begin{align}\label{pi0dim}
\log \dim (\Pi_0)\leq (\mumax&+\Delta)\log(\mumax+\Delta+m_k-1)\nonumber\\
&+\log(\mumax-\mumin+2\Delta+1).
\end{align}
%The first inequality in \eqref{typrob} is a Chernoff bound and holds for all $\Delta$, and the second follows from appropriate choice of $\Delta$. In this rewriting we have corrected some minor errors in the statement given in \cite{PhysRevA.89.062305}.
We choose any $\epsilon'>0$ (which can be optimized over), and fix $\Delta$ such that $\bra{\psi_{xy}}\Pi_0\ket{\psi_{xy}}\geq 1-\epsilon'$.

 We now bound the quantities $\tH(\Gamma^{AB})$, $\gamma$, and $\dim(AB)$ for a given choice of $\epsilon'$. First, by the dimension bound on entropy, $\tH(\Gamma^{AB})$ is upper bounded by Eq. \eqref{pi0dim}. Second,
\begin{align}\label{tdbound}
%&\left\Vert \rho^{AB}-\sigma \right\Vert_1\nonumber\\
%&\hspace{.2in}
\gamma & \leq \frac{1}{2} \sum_{x,y\in\{0,1\}^n} P(x,y) \left\Vert \ketbra{\psi_{xy}}{\psi_{xy}}-\frac{\Pi_0 \ketbra{\psi_{xy}}{\psi_{xy}} \Pi_0}{\text{Tr}(\Pi_0 \rho^{AB})}\right\Vert_1 \nonumber\\
&\leq \sum_{x,y\in\{0,1\}^n} P(x,y) \sqrt{1-\abs{\bra{\psi_{xy}}\Pi_0\ket{\psi_{xy}}}^2}\nonumber\\
&\leq \sqrt{1-(1-\epsilon')^2}\nonumber\\
&< \sqrt{2\epsilon'}
\end{align}
where the first inequality is the triangle inequality, the second is the Fuchs-van de Graaf inequality along with $\text{Tr}(\Pi_0 \rho^{AB})\leq 1$, the third is our parameter choice ensuring that $\bra{\psi_{xy}}\Pi_0\ket{\psi_{xy}}\geq 1-\epsilon'$, and the fourth is straightforward.
Third, as there are $2^{2n}$ states $\kett{\psi_{xy}}$, they span at most a $2^{2n}$-dimensional space. Combining these bounds, the Fannes-Audenaert inequality gives
\begin{align}\label{finitebound}
%\text{H}(\rho^{AB})&\leq \text{H}\left(\sigma\right)+\log(|X||Y|) \left\Vert \rho^{AB}-\sigma \right\Vert_1+ \text{h}\left(\frac{1}{2}  \left\Vert \rho^{AB}-\sigma \right\Vert_1\right)\nonumber\\
%\text{H}(\rho^{AB})&\leq \text{H}\left(\sigma\right)+\log(|X||Y|) \sqrt{2 \epsilon'}+\text{h}(\sqrt{2\epsilon'})\nonumber\\
\text{H}(\rho^{AB})&\leq \log\dim(\Pi_0)+2n \sqrt{2 \epsilon'}+\text{h}(\sqrt{2\epsilon'}).
\end{align}
The quantity \eqref{finitebound} holds for any distribution $P$, and thus upper bounds the information leakage. Although this bound is easily calculable in practice, it is linear in $n$. For all optical protocols we consider, $n$ is linear in $m_k$, so this bound does not give the desired $\mathcal{O}(\log m_k)$ asymptotic behaviour.

\subsection{Information leakage asymptotic analysis}\label{asymptoptical}
Here we prove the $\mathcal{O}(\log m_k)$ asymptotic information leakage of any simultaneous message passing model protocol satisfying the conditions outlined above. By Eq. \eqref{infleak}, the information leakage is equal to the entropy of $\rho^{AB}$, maximized over prior distributions $P\in \Pr(X\times Y)$. We bound $\tH(\rho^{AB})$ as $\mathcal{O}(\log m_k)$ by projecting onto Fock states lying within telescoping neighbourhoods $\Delta_0,\Delta_1,\dots$ of $[\mumin, \mumax]$.

In general, consider any projective measurement $\{\Pi_0,\Pi_1\dots,\}$ (with possibly infinitely many measurement operators), and define an isometry
\begin{align}
V=\sum_{j=0}^\infty \Pi_i \otimes \ket{j} \otimes \ket{j} \in \mathcal{U}(AB,AB C_1 C_2).
\end{align}
Then,
\begin{align}
\text{H}(AB)_{\rho}&= \text{H}(AB C_1 C_2)_{V\rho V^{\dagger}}\nonumber\\
&= \text{H}(C_1)+\text{H}(AB | C_1C_2)\nonumber\\
&\leq \text{H}(C_1)+\text{H}(AB|C_1)\nonumber\\
&\leq \text{H}(C_1)+\sum_{j=0}^\infty \text{Pr}(C_1=j) \log\dim (\Pi_j),\label{cohdim1}
\end{align}
where the first equality follows from the fact that isometries preserve entropy and the second equality follows from the chain rule and $\text{H}(C_1C_2)=\text{H}(C_1)$. The first inequality follows from strong subadditivity, and the second inequality follows from the dimension bound on quantum entropy.

%We choose a projective measurement onto telescoping neighbourhoods $\Delta_0,\Delta_1,\dots$ of $[\mumin, \mumax]$ such that $\Pr(E_1=i)$ decreases exponentially with $i$ and $\log\dim (\Pi_i)$ is $\mathcal{O}(\log n)$ (with prefactors not growing too quickly with $i$) to bound \eqref{cohdim1} as $\mathcal{O}(\log n)$.
For a fixed positive integer $\Delta$, let
\begin{align}
\Delta_0&=\{N\geq0 : \mumin-\Delta \leq N \leq \mumax+\Delta\} \nonumber\\
\Delta_j&=\{N\geq0: j \Delta+1 \leq \mumin-N\leq (j+1)\Delta \nonumber\\
&\hspace{.6in} \text{ or } \hspace{.1in}j \Delta+1 \leq N-\mumax \leq (j+1)\Delta\}
\end{align}
for each $j=0,1,\dots$. Let $\Pi_j$ be the projection onto the space of Fock states with total photon number lying in the set $\Delta_j$. Then the set $\{\Pi_0,\Pi_1,\dots\}$ forms a measurement.

We now show that $\Pr(C_1=j)$ decreases exponentially with $j$ and $\log\dim (\Pi_j)$ is $\mathcal{O}(\log m_k)$ for each $j$ (with prefactors not growing too quickly with $j$) to bound Eq. \eqref{cohdim1} as $\mathcal{O}(\log m_k)$. Using similar techniques to those used to prove Theorem 1 of \cite{PhysRevA.89.062305}, it can be shown that
\begin{align}\label{opt_chern}
\text{Pr}(C_1=j)&\leq e^{-\mumax}\left(\frac{e \mumax}{\mumax+j\Delta}\right)^{\mumax+j\Delta}
\end{align}
 for all $j=0,1,\dots$ under the simplifying assumption $\Delta>\mumax$, and
\begin{flalign}\label{opt_dim}
&\log\dim (\Pi_j) \leq (\mumax+(j+1)\Delta)\log(\mumax+(j+1)\Delta+m_{k}-1)\nonumber\\
&\hspace{1in}+\log(\Delta) \hspace{.3 in} \text{for all $j=1,2,\dots$}
\end{flalign}
It is straightforward to show that under these bounds the infinite sum appearing in the second term of Eq.  \eqref{cohdim1} converges and is $\mathcal{O}(\log m_{k})$.

It is also straightforward to show that $\text{H}(C_1)$ is no greater than the entropy of the $\mumax$ Poisson distribution, which is finite and constant in $m_k$ (in fact, it is well-approximated by $\frac{1}{2} \log( 2\pi e \mumax)$ when $\mumax \gg 1$ \cite{doi:10.1137/1030059}). Thus, the asymptotic information leakage is $\mathcal{O}(\log m_{k})$.

%It is straightforward to show that $\text{H}(C_1)$ is no greater than the entropy of the $\mumax$ Poisson distribution, which is finite and is well-approximated by $\frac{1}{2} \log( 2\pi e \mumax)$ when $\mumax \gg 1$ \cite{doi:10.1137/1030059}. Now we treat the second term of \eqref{cohdim1}. We make the choice $\Delta= \cl{\mumax}$ because it simplifies the asymptotic analysis. Under this choice, by a Chernoff bound,
%\begin{align}\label{opt_chern}
%\text{Pr}(E_1=i)&\leq e^{-\mumax}\left(\frac{e \mumax}{\mumax+i\Delta}\right)^{\mumax+i\Delta}
%\end{align}
%for all $i=0,1,\dots$ Note that \eqref{opt_chern} decreases exponentially with $i$.  Using the same technique as was used to prove Theorem 1 of \cite{PhysRevA.89.062305} it can be shown that
%\begin{align}\label{opt_dim}
%&\log\dim (\Pi_i) \leq (\mumax+(i+1)\Delta)\log(\mumax+(i+1)\Delta+m_{k}-1)\nonumber\\
%&\hspace{1in}+\log(\Delta) \hspace{.3 in} \text{for all $i=1,2,\dots$}.
%\end{align}
%Using the bounds \eqref{pi0dim}, \eqref{opt_chern} and \eqref{opt_dim} it is straightforward to show that the second term of \eqref{cohdim1} is $\mathcal{O}(\log m_{k})$. For every protocol that we consider, $m_{k}$ is linear in $n$, so the information leakage is $\mathcal{O}(\log n)$.

\bibliography{uw-ethesis}

\end{document}